\documentclass[aip,cha,reprint,amsmath,amssymb,floatfix]{revtex4-1}

\usepackage[utf8]{inputenc}
\usepackage[T1]{fontenc}
\usepackage{mathptmx}
\usepackage{bm}
\usepackage{graphicx}
\usepackage{xcolor}
\usepackage{etoolbox}
\usepackage{hyperref}
\hypersetup{hidelinks}

\makeatletter
\long\def\present@FM@footnote#1#2{%
 \begingroup
  \csname c@\@mpfn\endcsname#1\relax
  \def\@thefnmark{\frontmatter@thefootnote}%
  \par\addvspace{5\p@}%
  \normalfont\small
  \noindent
  \Hy@raisedlink{\hyper@anchorstart{frontmatter.#1}\hyper@anchorend}%
  \frontmatter@makefnmark #2\par
 \endgroup
}%
\makeatother

\renewcommand{\vec}{\bm}
\newcommand{\mat}{\bm}

\begin{document}


\title{Intermittent flocking and fractal collective order induced by time-varying delays}


\author{David Müller-Bender}
\email[]{david.mueller-bender@mailbox.org}
\affiliation{Institute of Physics, Chemnitz University of Technology, 09107 Chemnitz, Germany}
\author{Rahil N. Valani}
\email[]{rahil.valani@physics.ox.ac.uk}
\affiliation{Rudolf Peierls Centre for Theoretical Physics, Parks Road, University of Oxford, OX1 3PU, United Kingdom}



\date{\today}

\begin{abstract}
Time-varying interaction delays are ubiquitous in active matter, yet their collective effects remain largely unexplored. We show that active particles with internal dynamics and Vicsek-like delayed alignment exhibit \emph{intermittent flocking}, characterized by long episodes of coherent motion interrupted by brief disordering events. This collective behavior arises from \emph{laminar chaos}, a form of chaotic dynamics unique to systems with time-varying delays. Changing only the delay parameters qualitatively reshapes collective motion, producing fractal changes in global flocking order. Our results establish the temporal structure of interaction delays as a new control parameter for active matter.
\end{abstract}

\maketitle

\begin{quotation}
Moving together requires responding to others, but those responses are rarely instantaneous. Here, we show that a temporal variation of the response time can fundamentally reshape collective motion. In a model of self-propelled particles, time-varying delays generate \emph{intermittent flocking}: long episodes of coordinated motion interrupted by brief losses of order. This behavior originates from \emph{laminar chaos}, in which the direction of motion remains nearly constant for long intervals between irregular changes. Disorder then helps restore order: particles moving in different directions encounter new neighbors, allowing the flock to realign. This feedback sustains the repeated loss and recovery of collective order. Varying the delay parameters also produces a fractal pattern, with regions of stronger and weaker collective order interwoven across scales. Our results connect the dynamics of delayed responses to collective motion and identify the temporal structure of interaction delays as a means of shaping how groups move together.
\end{quotation}

\section{Introduction} Active matter systems, from biological collectives to synthetic self-propelled particles, exhibit a wealth of emergent phenomena, including motility-induced phase separation~\cite{Cates2015}, active turbulence~\cite{Alert2022}, and flocking~\cite{Vicsek1995Phase,Vicsek2012Collective}. Flocking, the spontaneous alignment of particle velocities into coherent collective motion, is the paradigmatic example of self-organization in active matter. The seminal Vicsek model established the minimal mechanism for the emergence of collective motion~\cite{Vicsek1995Phase}, while subsequent extensions incorporated ingredients such as noise, heterogeneity, confinement, and non-metric interactions~\cite{Vicsek2012Collective}. More recently, attention has turned to active matter models with internal degrees of freedom~\citep{frasca_synchronization_2008,Valaniattractormatter2023,SAR20261}. In particular, {swarmalator} models demonstrate that coupling self-propulsion to internal phase dynamics gives rise to synchronized clusters, phase waves, and chimera-like states, revealing how internal dynamics can fundamentally reshape collective motion~\cite{OKeeffe2017,SAR20261}.

\begin{figure}
    \centering
\includegraphics[width=\linewidth]{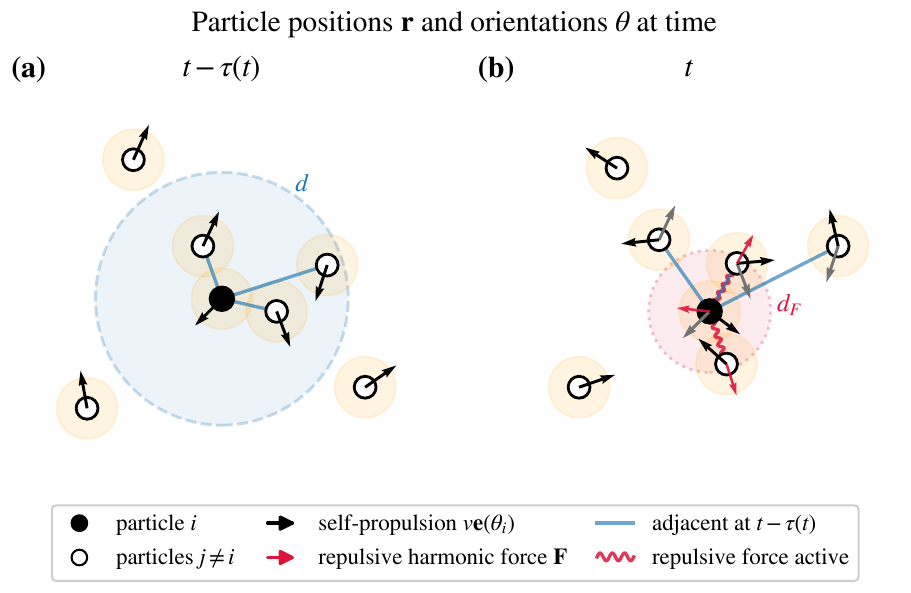}

\caption{
	Schematic of the model. All particles self-propel at constant speed $v$, where the directions are indicated by the black arrows. \textbf{(a)} Delayed particle positions at time $t-\tau(t)$ determine the neighborhood of active particle $i$, where particles inside the interaction radius $d$ (blue disk) are adjacent. \textbf{(b)} The same adjacency links connect the particles at their new positions at time $t$. The orientation $\theta_i(t)$ of particle $i$ evolves according to the delayed alignment dynamics in Eq.~\eqref{eq:heading_dde}, coupling its orientation to the delayed orientations $\theta_{j}(t-\tau(t))$ (gray arrows, same direction as in (a)). Particles are modeled as soft disks (small yellow disks) and interact through an  instantaneous short-range harmonic repulsion acting for separations smaller than $d_F$ (red disk). The red arrow at particle $i$ shows the sum of repulsive forces from surrounding particles, whereas the two other represent the repulsive forces that the respective particles experience from particle~$i$.
	}
    \label{fig:model}
\end{figure}

A second ingredient receiving increasing attention is the role of time delays. Delays arise naturally from finite sensing, signal propagation, information processing, and memory, and are known to induce new collective states, oscillations, pattern formation, and phase transitions in active matter~\cite{PhysRevE.106.054612,PhysRevLett.127.258001,Horton_2025,PAKPOUR2024129453,PhysRevE.90.062708,Costanzo_2022,PhysRevE.93.032307,Chen2011,5400490,11489308,10.1145/3760269.3760339,li_informational_2026}. In many biological and synthetic systems, however, interaction delays are themselves time dependent, reflecting dynamically evolving sensing and processing times. Yet most previous studies have focused on constant delays or their average magnitude, leaving open the question of whether the temporal structure of interaction delays can itself influence collective behavior.

Recent advances in delay-dynamical systems provide a natural framework for addressing this question. In particular, they reveal a distinct form of dynamics, termed \emph{laminar chaos}, that arises uniquely from time-varying delays~\cite{muller_laminar_2018}.
Unlike conventional chaos, laminar chaos is characterized by a time series with long, nearly constant episodes that are repeatedly interrupted by brief irregular bursts, giving rise to intrinsically multiscale dynamics. Although well understood in low-dimensional delay systems, its consequences for spatially extended active matter remain essentially unexplored.

These observations raise a natural question: \emph{Can the temporal structure of interaction delays fundamentally reshape collective motion in active matter?}

In this work, we show that the temporal structure of interaction delays fundamentally changes flocking dynamics. Active particles with internal states exhibit \emph{intermittent flocking}, in which long episodes of coherent motion are interrupted by brief disordering events. The underlying mechanism is an exact reduction of the fully ordered flock to a scalar delay equation exhibiting laminar chaos. As the delay parameters are varied, the system exhibits fractal changes in collective order, revealing time-varying delays as a new organizing principle for collective motion.

\section{Model} We consider $N$ active particles moving inside a two-dimensional unit square domain with periodic boundary conditions as illustrated in Fig.~\ref{fig:model}. The dynamics of the $i$th active particle at position $\vec{r}_i=(x_i,y_i)^\top$ is given by
\begin{equation}
	\label{eq:position_ode}
	\dot{\vec{r}}_i(t)
	=
	v
	\begin{pmatrix}
		\cos\theta_i(t)\\
		\sin\theta_i(t)
	\end{pmatrix}
	+
	\sum_{j\neq i}\vec{F}(\vec{r}_i(t),\vec{r}_j(t)),
	\quad i=1,\dots,N ,
\end{equation}
where $\vec{F}$ is a repulsive force between particle pairs.
The internal dynamics that determines the particle orientation $\theta_i$ is a delayed differential equation (DDE) with Vicsek-like alignment rule and a finite turning rate $T$, given by
\begin{equation}
	\label{eq:heading_dde}
	\frac{1}{T}\dot\theta_i(t)+\theta_i(t)
	=
	f(\theta_i(R(t)))
	-
	\sigma
	\sum_{j=1}^N
	L_{ij}(R(t)) f(\theta_j(R(t))).
\end{equation}
Its structure resembles the internal dynamics of mobile oscillators~\citep{fujiwara_synchronization_2016} and biological network motifs, where $f$ represents activation or repression and $1/T$ sets the relaxation time scale \cite{glass_nonlinear_2021}. For an uncoupled particle ($\sigma=0$), Eq.~\eqref{eq:heading_dde} reduces to a nonlinear delayed-feedback oscillator, a paradigmatic model in physiology \cite{mackey_oscillation_1977}, optics \cite{ikeda_multiple-valued_1979,ikeda_optical_1980}, and optoelectronics \cite{hart_laminar_2019,larger_complexity_2013,chembo_optoelectronic_2019}. Here, we extend networks of such delayed oscillators \cite{sysoev_recovery_2016,ponomarenko_chimeralike_2017,chembo_optoelectronic_2019} to interacting active particles.

The first term describes the intrinsic orientation dynamics, where $R(t)=t-\tau(t)$ is the retarded argument. The delay $\tau(t)$ models finite sensing and information-processing times, while the nonlinear response function $f$ determines the preferred heading from delayed orientational information. The second term describes Vicsek-like alignment with neighbors within an interaction radius $d$, defining the time-dependent interaction network with adjacency matrix
\begin{equation}
	A_{ij}(t)
	=
	\begin{cases}
		1, & i\neq j \text{ and } \|\vec{r}_i(t)-\vec{r}_j(t)\|_{\mathbb T}<d,\\
		0, & \text{otherwise},
	\end{cases}
\end{equation}
and the normalized Laplacian $L_{ij}(t)$ given by
\begin{equation}
	L_{ij}(t)
	:=
	\begin{cases}
		\displaystyle \delta_{ij}-\frac{A_{ij}(t)}{k_i(t)}, & k_i(t)>0,\\[2mm]
		0, & k_i(t)=0,
	\end{cases}
\end{equation}
where $k_i(t) = \sum_{j=1}^N A_{ij}(t)$ is the instantaneous number of neighbors of particle $i$ and $\|\cdot\|_{\mathbb T}$ denotes the euclidean torus distance.
To suppress particle accumulation, we introduce a short-range harmonic repulsive force $\vec{F}(\vec{r}_i,\vec{r}_j) = -\kappa\left(d_F-\|\vec{r}_i-\vec{r}_j\|_{\mathbb T}\right) \vec{n}_{ij}$, acting within a distance $d_F$, where $\kappa$ is the repulsion strength and $\vec{n}_{ij}$ is the unit vector pointing from particle $i$ to particle $j$ along the shortest path on the torus.

For simplicity we consider the periodically varying delay
\begin{equation}
\label{eq:delay}
\tau(t)=\tau_0+\frac{A}{2\pi}\sin(2\pi t),
\end{equation}
as done in the original theory of laminar chaos \cite{muller_laminar_2018}, while noting that laminar chaos also occurs for quasiperiodic \cite{muller-bender_laminar_2023}, random, and chaotic delays \cite{muller-bender_laminar_2025}.
Although these cases incorporate interesting new details, the conditions for laminar chaos and its fundamental properties persist.
The results presented here therefore can be generalized to more complex delay variations. 
For the nonlinear response function we choose
\begin{equation}
\label{eq:nonlinearity}
f(\theta)=\theta+a\sin(\theta),
\end{equation}
with $a=2\pi\times0.9$, for which an isolated particle ($\sigma=0$) exhibits chaotic orientational diffusion \cite{albers_chaotic_2022}. Our results require chaotic internal dynamics but are not specific to this choice of $f$. Unless stated otherwise, we use $N=1000$, $v=10$, $T=10^4$, $\sigma=1$, $d=0.076$, $\kappa=1$, and $d_F=d/2$. These parameters place the system in the regime of large $N$, $v$, and $T$, while $d$ is chosen such that particles interact with a fixed average number of neighbors.

\begin{figure}
    \centering
    \includegraphics[width=1\linewidth]{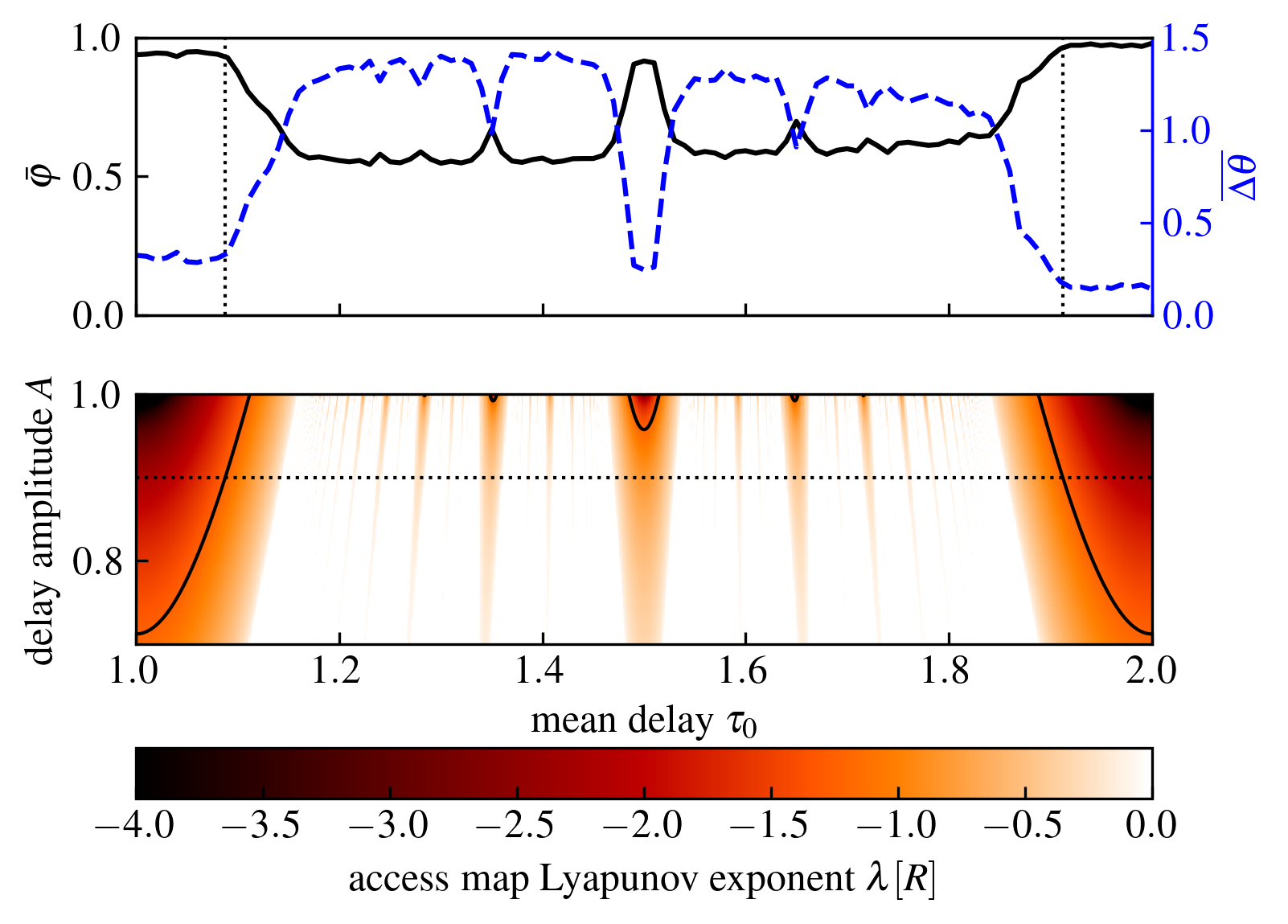}
\caption{Fractal organization of collective order on the delay parameters. \textbf{Top:} Time-averaged flocking order parameter $\bar{\varphi}$ (solid) and time-averaged synchronization error $\overline{\Delta\theta}$ (dashed) as functions of the mean delay $\tau_0$ for fixed delay amplitude $A=0.9$. The collective order varies fractally with the delay parameters. \textbf{Bottom:} Lyapunov exponent $\lambda[R]$ of the access map $t'=R(t)=t-\tau(t)$, which classifies time-varying delays into conservative ($\lambda[R]=0$) and dissipative ($\lambda[R]<0$) regimes. The Arnold tongues ($\lambda[R]<0$) coincide with regions of enhanced flocking order in the top panel. The solid contour marks the onset of classical laminar chaos, $\lambda[f]+\lambda[R]=0$, while the dotted horizontal lines indicate the parameter values shown in the top panel.}
    \label{fig:order_parameter}
\end{figure}

\begin{figure}[!t]
    \centering
    \includegraphics[width=0.9\linewidth]{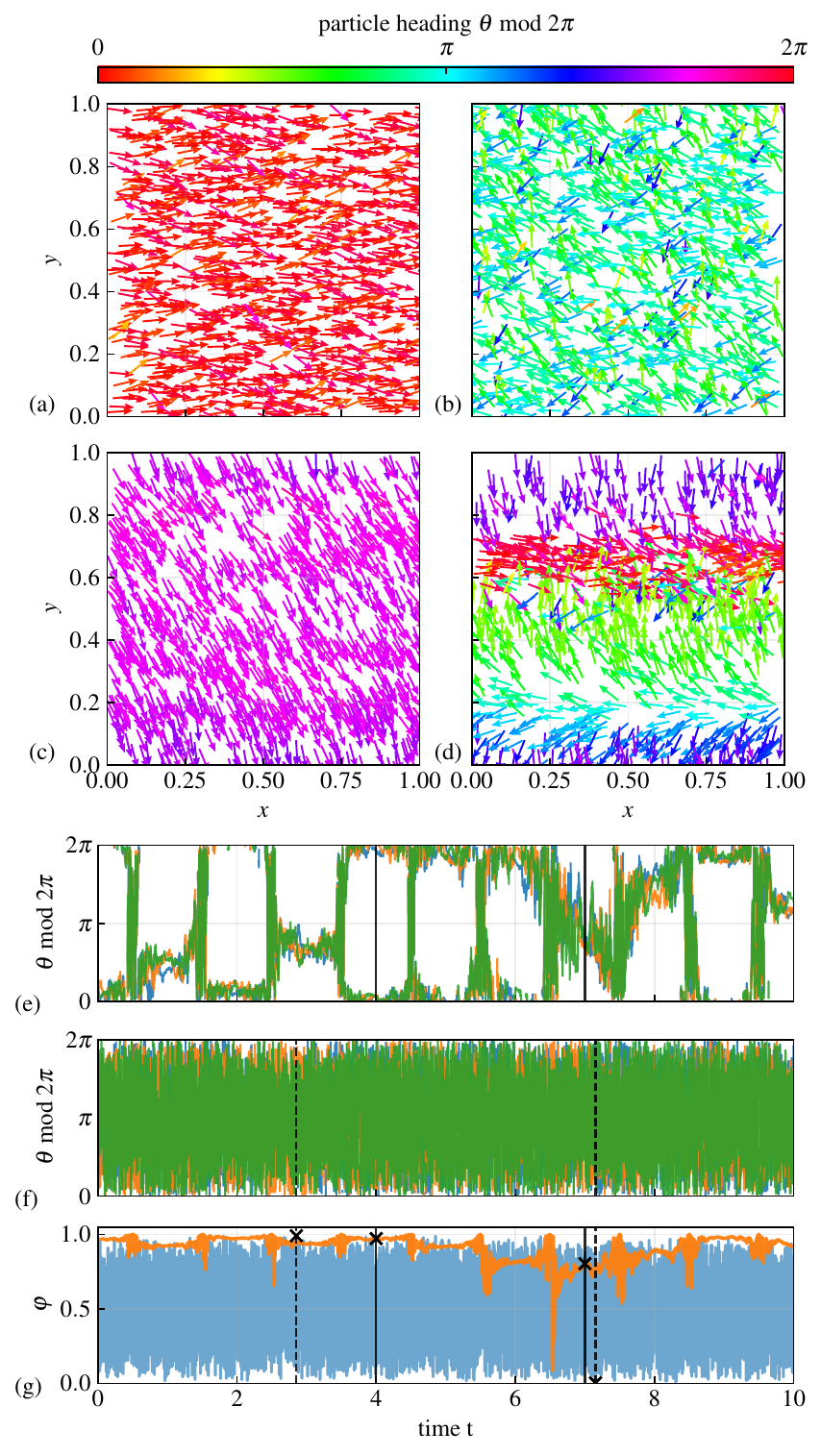}
\caption{Collective regimes induced by time-varying delays. Arrows and colors indicate particle orientations. \textbf{(a,b)} Dissipative delays with $\lambda[f]+\lambda[R]<0$ ($A=0.9$, $\tau_0=1$) yield intermittent flocking: ordered laminar phases (a) and intra-plateau disordering events (b). \textbf{(c,d)} Conservative delays ($A=0$, $\tau_0=1$) give rise to rapidly fluctuating collective dynamics, alternating between highly ordered (c) and weakly ordered (d) flocking states. \textbf{(e,f)} Headings of three representative particles for the dissipative and conservative regimes, respectively; vertical lines mark snapshot times for (a-d). \textbf{(g)} Flocking order for dissipative and conservative delays with values at snapshot times (crosses). See Supplementary Videos 1 and 2.
}
    \label{fig:timeseries}
\end{figure}

\section{Results}
We begin by exploring how the collective dynamics depend on the delay parameters, namely the delay amplitude $A$ and mean delay $\tau_0$ \footnote{For the numerical results, Eqs.~\eqref{eq:position_ode} and \eqref{eq:heading_dde} were solved using the trapezoidal rule and the Lobatto IIIC method \cite{bellen_numerical_2003}, respectively, where linear interpolation was used for the delayed term. We used a solver time step $\Delta t = (10\,T)^{-1}$ and initialized the system with random constant initial history functions uniformly distributed in $[0,1]\times [0,1]$ for the position and $[0,2\pi]$ for the heading. Before computing the observables, we let the transients relax for at least $1000$ delay periods. Drafting of the source code was assisted by ChatGPT 5.3-5.5, where the code was reviewed and tested manually.}. Remarkably, the system exhibits a fractal dependence of the collective order on the delay parameters, as shown in the top panel of Fig.~\ref{fig:order_parameter}. For fixed $A$ and varying $\tau_0$, we compute the time-averaged flocking order parameter, 
$\bar{\varphi}
=
(t_\mathrm{end}-t_0)^{-1}
\int_{t_0}^{t_\mathrm{end}}
dt\,\varphi(t)$, 
together with the time-averaged synchronization error $\overline{\Delta\theta}$. The flocking order parameter, $\varphi(t)=\left|\left\langle e^{\mathrm{i}\theta_i(t)}\right\rangle_i\right|$, measures the global polar order of the flock. Since flocking corresponds to global alignment of particle orientations, the flocking order parameter is equivalent to the Kuramoto order parameter. The synchronization error is defined as
$\Delta\theta(t)
=
\sqrt{\left\langle
\left[\theta_i(t)-\langle\theta_i(t)\rangle_i\right]^2
\right\rangle_i}$,
where $\langle\cdot\rangle_i$ denotes the ensemble average over all particles \footnote{The time averages were computed from time series of length $t_\mathrm{end}-t_0=1000$ after discarding an initial transient of duration $t_0=1000$. The time step used for the averaging was $10^{-2}$, while the numerical integration time step was $10^{-5}$.}.

The origin of the fractal dependence is revealed in the bottom panel of Fig.~\ref{fig:order_parameter}, which shows the Lyapunov exponent $\lambda[R]$ of the one-dimensional \emph{access map},
$
t'=R(t)=t-\tau(t)$,
a key object governing the dynamics of systems with time-varying delays \cite{otto_universal_2017}. Since the access map is the lift of a circle map (cf.~\onlinecite{katok_introduction_1997}), the delay parameter space naturally separates into two classes. The Arnold tongues \cite{arnold_small_1961,*arnold_small_1961_erratum}, where $\lambda[R]<0$, correspond to \emph{dissipative delays}, whereas their complement, characterized by $\lambda[R]=0$, defines \emph{conservative delays}. This classification results in a fat fractal \cite{ott_chaos_2002} in the delay parameter space, which is directly reflected in the fractal dependence of the collective order shown in the top panel. Thus, the geometry of the delay dynamics is directly imprinted onto the collective order.
For the uncoupled system $(\sigma=0)$ we know that the delay classes give rise to fundamentally different types of chaos for large $T$ \cite{muller_laminar_2018,muller-bender_resonant_2019}.
In the limit $T\to\infty$ each component of Eq.~\eqref{eq:heading_dde} take the structure $\theta(t)=f(\theta(R(t)))$, which can be solved via the method of steps \cite{bellman_computational_1965} by introducing segments $\theta_{(l)}(t)$ with $t\in(R(t_l),t_l]$ giving $\theta_{(l)}(t)=f(\theta_{(l-1)}(R(t)))=f^l(\theta_{(0)}(R^l(t)))$, where power to $l$ means $l$-times functional iteration.
Therefore, a competition between two one-dimensional maps defined by $f$ and $R$ governs the dynamics, where for chaotic dynamics the Lyapunov exponent $\lambda[f]$ of the map defined by $f$ must be positive.
For conservative delays, $\lambda[R]=0$, the access map cannot compensate the high-frequent chaotic fluctuations that are induced by $f$ that increase from segment to segment until they are bounded by the relaxation rate $T$ if it is finite.
This leads to a high-dimensional type of chaos called \emph{turbulent chaos} \cite{ikeda_optical_1980,ikeda_high-dimensional_1987}.
In contrast, dissipative delays, $\lambda[R]<0$, lead to low-dimensional chaos, where \emph{classical laminar chaos} is observed if $\lambda[f]+\lambda[R]<0$ holds.
The latter is characterized by nearly constant laminar phases, where the durations are large compared to the relaxation time $(1/T)$ and the levels vary chaotically from phase to phase.
The remaining dissipative delays leads to \emph{(generalized) laminar chaos}, which is also low-dimensional compared to turbulent chaos but now polynomials take the role of the constant laminar phases \cite{muller-bender_resonant_2019}.
For brevity, we focus on classical laminar chaos which is observed in Fig.~\ref{fig:order_parameter}, where the delay amplitude $A$ is above the black contour line.

The two classes of delay give rise to qualitatively different forms of collective motion, illustrated in Fig.~\ref{fig:timeseries}. Conservative delays, including the case of constant delay, produce rapid fluctuations between highly ordered and weakly ordered flocking states, as reflected by the strongly fluctuating flocking order parameter (blue curve in Fig.~\ref{fig:timeseries}(g)). The corresponding particle snapshots [Figs.~\ref{fig:timeseries}(c,d)] reveal repeated transitions between coherent and disordered configurations, while the particle orientations fluctuate irregularly over time [Fig.~\ref{fig:timeseries}(f)]. This behavior originates from the turbulent-chaotic dynamics of the underlying uncoupled delay system ($\sigma=0$)~\citep{muller_laminar_2018} and resembles collective chaos, in which macroscopic fluctuations persist in the thermodynamic limit \cite{shibata_collective_1998,cencini_macroscopic_1999,pazo_quasiperiodic_2016}.

Dissipative delays, by contrast, give rise to a qualitatively different form of collective motion: \emph{intermittent flocking}. Rather than fluctuating continuously, the flock exhibits long-lived episodes of nearly complete polar order that are interrupted by brief disordering events [orange curve in Fig.~\ref{fig:timeseries}(g)]. During the laminar phases, particles move with almost identical orientations [Fig.~\ref{fig:timeseries}(a)], as reflected in the heading dynamics [Fig.~\ref{fig:timeseries}(e)]. This behavior originates from the laminar-chaotic dynamics of the corresponding uncoupled scalar system ($\sigma=0$), which arises whenever $\lambda[f]+\lambda[R]<0$. The brief disordering events correspond to transitions between successive laminar plateaus, while the rare intra-plateau disordering events reflect residual fluctuations within a laminar phase.

To explain the origin of intermittent flocking, we analyze the stability of the fully ordered flock, $\theta_1(t)=\theta_2(t)=\cdots=\theta_N(t)=:\theta_s(t)$
for which all particles move with a common orientation \footnote{The analytical derivations were assisted by ChatGPT 5.5, where all derivations and resulting formulas were verified manually.}. Crucially, on this synchronization manifold, the collective dynamics reduce exactly to the scalar delay equation
\begin{equation}
	\frac{1}{T}\dot\theta_s(t)+\theta_s(t) = f(\theta_s(R(t))).
\end{equation}
Thus, the low-dimensional dynamics governing laminar and turbulent chaos~\citep{muller_laminar_2018} is embedded exactly into the high-dimensional flocking dynamics, explaining why changes in the delay dynamics directly reshape collective motion. To study the stability of this collective state, we introduce small perturbations $\vec{\xi}(t)=(\xi_1,\xi_2,\ldots,\xi_N)^\top$ of the particle orientations. Linearization of Eq.~\eqref{eq:heading_dde} around this state yields
\begin{equation}
	\label{eq:lin:heading}
	\frac{1}{T}\dot{\vec{\xi}}(t)+\vec{\xi}(t) = f'(\theta_s(R(t))) \left[ \mat{I}-\sigma \mat{L}^s(R(t))	\right]	\vec{\xi}(R(t)),
\end{equation}
where $\mat L^s(t)$ denotes the interaction network evaluated along the synchronized trajectory.

The remaining challenge is the time-dependent interaction network. Along the synchronized trajectory, the particle configuration is stationary up to a global translation, so the network is effectively constant. Away from synchrony, however, its dynamics depends on whether the synchronized state is turbulent or laminar. Turbulent-chaotic orientations fluctuate on the short time scale $1/T$, producing only weak particle diffusion over a delay time, with diffusion coefficient proportional to $1/T$ (see Appendix~\ref{app:particle-mobility}). The network therefore remains sparse for small interaction radius $d$. Laminar-chaotic orientations, by contrast, contain long nearly constant plateaus that produce persistent particle motion. For sufficiently large propulsion speed $v$, this rapidly reshuffles neighbors and efficiently mixes the network.
Following Refs.~\onlinecite{frasca_synchronization_2008,fujiwara_synchronization_2016}, we approximate the rapidly reshuffling interaction network by its time average using the fast-switching approximation \cite{belykh_blinking_2004,porfiri_random_2006,stilwell_sufficient_2006}, giving
$L^s_{ij}(t) \approx \langle L^s_{ij}(t) \rangle_t = \sigma_0\,[\delta_{ij} - (1-\delta_{ij})/(N-1)]$, where $\sigma_0=1-(1-\pi d^2)^{N-1}$ is the probability that a uniformly distributed particle has at least one neighbor and $d\leq1/2$.

The stability of the synchronized flock can be analyzed using the master-stability framework \cite{pecora_master_1998}. For a constant interaction Laplacian, $\mat{L}^s(t)=\mat{L}^*$, with eigenvalues $\mu_\alpha$, diagonalizing Eq.~\eqref{eq:lin:heading} decouples the perturbation dynamics into independent modes $\eta(t)$ governed by the scalar variational equation
\begin{equation}
\label{eq:master_stability_dde}
\frac{1}{T}\dot{\eta}(t)
=
-\eta(t)
+
f'(\theta_s(R(t)))
\left(1-\sigma\mu\right)
\eta(R(t)).
\end{equation}
The eigenmode with $\mu_0=0$ lies tangent to the synchronization manifold, while all modes with $\mu_\alpha>0$ are transverse and therefore determine the stability of the ordered flock. Stability is characterized by the master-stability function $\Lambda(\sigma\mu)$, defined as the largest Lyapunov exponent of Eq.~\eqref{eq:master_stability_dde}. The synchronized state is stable whenever $\Lambda(\sigma\mu_\alpha)<0$ for all transverse modes.

\begin{figure}
    \centering
    \includegraphics[width=0.9\linewidth]{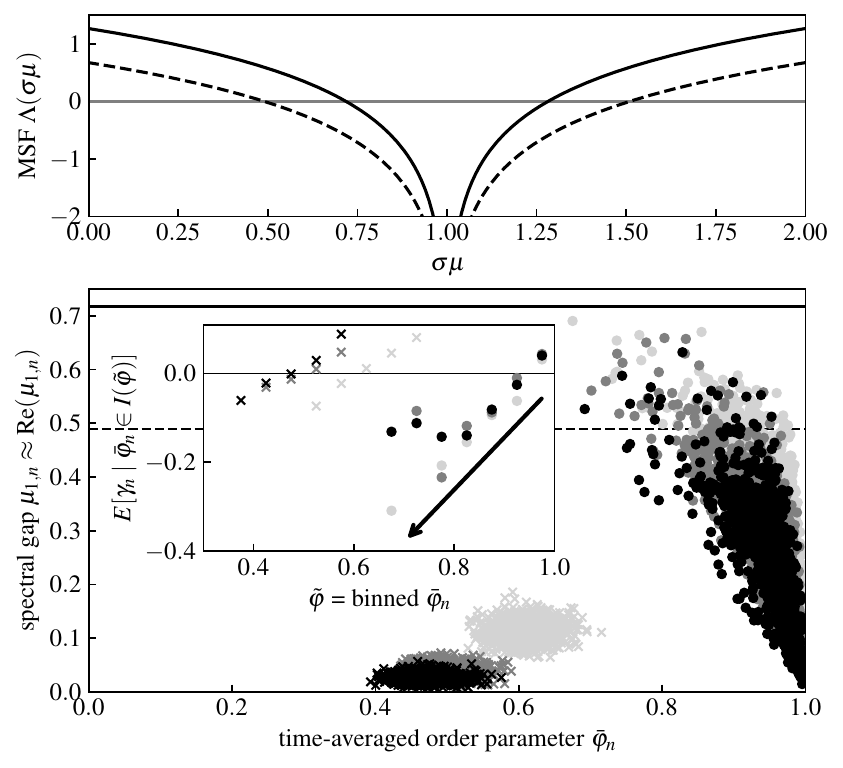}
\caption{Self-sustained intermittent flocking. \textbf{Top:} Master stability function (MSF) for laminar (solid, $A=0.9$, $\tau_0=1$) and turbulent chaos (dashed, $A=0$, $\tau_0=1$). \textbf{Bottom:} Spectral gap $\mu_{1,n} \approx \text{Re}(\mu_{1,n})$ of the interval-averaged network \cite{Note4} versus interval-averaged flocking order parameter $\bar{\varphi}_n$. During laminar chaos (dots) disorder reshuffles neighbors, therefore increases $\mu_{1,n}$ and restores order; turbulent chaos (crosses) leaves the network sparse. The solid (dashed) horizontal line marks the stability threshold $\Lambda(\sigma\mu_1)=0$ for laminar (turbulent) chaos. \textbf{Inset:} Conditional mean recovery rate $E[\gamma_n \mid \bar{\varphi}_n \in I(\tilde{\varphi})]$ becomes negative at low order. Results are shown for $N=200$, $600$, and $1000$ particles (light gray, gray, and black, respectively), with interaction radii $d=0.17$, $0.098$, and $0.076$ chosen to maintain the average number of neighbors, $N\pi d^2\approx18$.
}
    \label{fig:laplacian_gap_vs_sync}
\end{figure}

In the top panel of Fig.~\ref{fig:laplacian_gap_vs_sync}, we show the master-stability function for the laminar- and turbulent-chaotic regimes. Its minimum, $\Lambda(\sigma\mu=1)=-T$, is independent of the nonlinear function $f$ and corresponds to the idealized all-to-all coupling limit, where $\sigma=1$ and $\mu_{\alpha>0}=N/(N-1)\approx1$ for large $N$. In our system, however, the finite interaction radius and steric repulsion keep the interaction network sparse, even in a fully ordered flock. Consequently, for each fixed time $t$ the spectral gap of $\mat{L}^s(t)$ satisfies $\mu_1\approx0$, placing transverse modes in the unstable region, $\Lambda(\sigma\mu_1)>0$. However, if neighbors are reshuffled fast compared to the system dynamics, time-averages of $\mat{L}^s(t)$ and the corresponding spectral gaps have to be considered instead of instantaneous values as done below.

The subsequent evolution is therefore governed by how disorder modifies the interaction network. To quantify this feedback, we divide the dynamics into non-overlapping intervals of one delay length, $t\in[t_{n-1}=R(t_n),t_n]$, and compute the spectral gap $\mu_{1,n}$ of the interval-averaged Laplacian $\bar L_{ij,n}$, the synchronization recovery rate $\gamma_n=\ln(\overline{\Delta\theta}_{n+1}/\overline{\Delta\theta}_n)$, and the corresponding interval-averaged flocking order $\bar\varphi_n$ \cite{Note2}\textsuperscript{,}\footnote{The time-averaged Laplacian $\bar{\mat{L}}_n$ is defined as
$\bar{L}_{ij,n}=[\tau(t_n)]^{-1}\int_{t_{n-1}}^{t_n} dt\, L_{ij}(t)$ and $\overline{\Delta\theta}_n=[\tau(t_n)]^{-1}\int_{t_{n-1}}^{t_n} dt\, \Delta\theta(t)$. While the eigenvalues of the instantaneous Laplacian $\mat{L}(t)$ are real, we obtain complex eigenvalues for the $\bar{\mat{L}}_{n}$ with a maximum imaginary part of order $10^{-2}$ over the whole dataset. We approximate $\mu_{\alpha,n}\approx \text{Re}(\mu_{\alpha,n})$ and consider the MSF $\Lambda(\sigma\mu)$ only for real arguments since deviations from $\Lambda(\sigma\mu+\imath \delta\mu)$, $\delta\mu \in [-10^{-2},10^{-2}]$ are not visible over the plot range in Fig.~\ref{fig:laplacian_gap_vs_sync} (top panel). Significant deviations occur near $\sigma\mu=1$ but $\Lambda$ remains negative in this region.}.  The results are shown in the bottom panel of Fig.~\ref{fig:laplacian_gap_vs_sync}.

During turbulent chaos, particle diffusion over one delay length is weak, so neighbors change little and the interval-averaged network remains sparse, keeping $\mu_{1,n}$ well below the stability threshold. Laminar chaos, by contrast, produces long plateaus of nearly constant orientation. When collective order decreases, the resulting persistent motion rapidly reshuffles neighbors, increasing $\mu_{1,n}$ and pushing the transverse modes toward the stable region of the master-stability function, $\Lambda(\sigma\mu)<0$. The flock then reorders, as reflected by the synchronization recovery rate $\gamma_n$. As this quantity strongly fluctuates, we consider its conditional expectation $E[\gamma_n\mid\bar{\varphi}_n\in I(\tilde{\varphi})]$, which becomes strongly negative for values of $\tilde{\varphi}$ that are significantly lower than one. It was computed via $E[\gamma_n \mid \bar{\varphi}_n \in I(\tilde{\varphi})] \approx \left(\sum_{n,\bar{\varphi}_n\in I(\tilde{\varphi})} \gamma_n \right)\big/ |\{n \mid \bar{\varphi}_n \in I(\tilde{\varphi})\}|$, where $I(\tilde{\varphi})=[\tilde{\varphi}-\delta\tilde{\varphi}/2,\tilde{\varphi}+\delta\tilde{\varphi}/2)$ and $\delta\tilde{\varphi}=0.05$. As order is restored, the network gradually becomes sparse again, driving $\mu_{1,n}$ below the stability threshold and initiating the next disordering event. Intermittent flocking is therefore sustained by a self-organized feedback between transient disorder, neighbor reshuffling, and restored synchrony. The mechanism is robust to parameter changes, as demonstrated in Appendix~\ref{app:robustness}, where $d$ is varied.

\section{Conclusions}
We showed that time-varying interaction delays induce \emph{intermittent flocking}, with \emph{laminar chaos} providing the underlying dynamical mechanism. The resulting self-organized feedback between transient disorder, neighbor reshuffling, and restored synchrony sustains the intermittent collective state, while the underlying delay dynamics imprint their Arnold-tongue geometry onto collective motion, producing abrupt, fractal changes in global order.

Our predictions could be tested in synthetic active matter, where delay effects have already attracted considerable attention \cite{PhysRevE.93.032307,chen_persistent_2024}. A particularly direct test would be to vary the temporal modulation of the interaction delays while keeping their mean fixed, which we predict will induce intermittent flocking and abrupt changes in global order. The required nonlinear delayed-feedback dynamics could be realized in optoelectronic systems supporting arbitrary networks of nonlinear elements \cite{hart_experiments_2017} and oscillators with time-varying delay \cite{hart_delayed_2019,muller-bender_laminar_2020}.

More broadly, our work establishes a direct bridge between modern delay-dynamical systems theory and active matter, revealing the temporal structure of interaction delays as a new organizing principle for collective motion. We expect this perspective to stimulate the exploration of time-varying delays across a broad range of active matter systems.

\section*{Supplementary Material}
Supplementary Videos 1 and 2 illustrate the collective dynamics shown in Fig.~\ref{fig:timeseries}: intermittent flocking for a dissipative delay and rapid fluctuations between highly and weakly ordered states for a conservative delay. The videos show the evolving particle positions alongside time series of particle headings and the flocking order parameter.

\section*{Acknowledgments}
D.MB. gratefully acknowledges funding by the Deutsche Forschungsgemeinschaft (DFG, German Research Foundation) - 456546951.
R.V. acknowledges the support of the Leverhulme Trust [Grant No. LIP-2020-014] and the ERC Advanced Grant ActBio (funded as UKRI Frontier Research Grant EP/Y033981/1).
The authors used ChatGPT versions 5.3–5.5 to assist with code drafting, analytical derivations, exploratory discussions, and manuscript editing. All code, analytical derivations, numerical results, figures, and scientific conclusions were checked, tested, and validated by the authors.

\section*{Author Declarations}

\subsection*{Conflict of Interest}
The authors have no conflicts to disclose.

\subsection*{Author Contributions}
\textbf{David Müller-Bender:} Conceptualization (equal); Investigation (lead); Software (lead); Writing -- original draft (equal); Writing -- review \& editing (equal). \textbf{Rahil N. Valani:} Conceptualization (equal); Investigation (supporting); Writing -- original draft (equal); Writing -- review \& editing (equal)

\section*{Data Availability}
The data that support the findings of this study are available from the corresponding author upon reasonable request.

\appendix
\section{Particle mobility}
\label{app:particle-mobility}

\begin{figure}[h]
    \centering
    \includegraphics[width=1\linewidth]{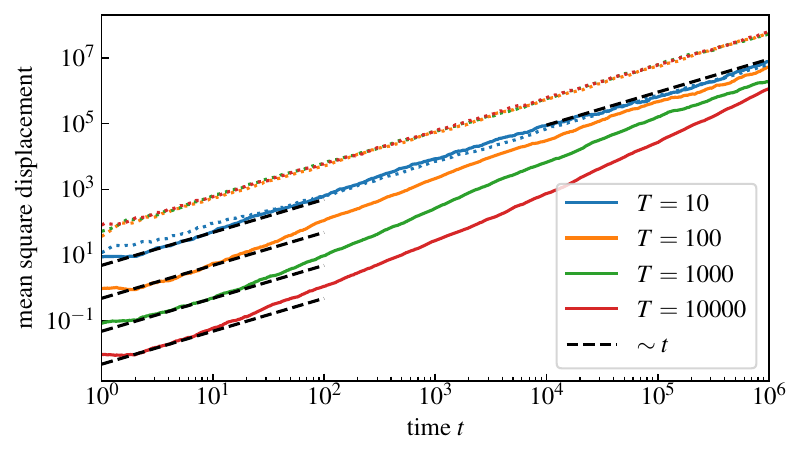}
    \caption{Particle mobility for turbulent and laminar chaos. Mean-square displacement for Eq.~\eqref{eq:heading_dde} with turbulent ($A=0$, $\tau_0=1$, solid lines) and laminar ($A=0.9$, $\tau_0=1$, dotted lines) chaos in the absence of coupling ($\sigma=0$) and repulsive interactions ($\vec{F}=\vec{0}$), after subtracting the common drift. The dashed lines indicate short-time diffusion coefficients $D_{\mathrm{short}}\approx50/T$ for turbulent chaos. By contrast, the diffusion coefficient for laminar chaos approaches a finite value as $T$ increases, indicating persistent motion and efficient mixing of the interaction network.}
    \label{fig:msd}
\end{figure}

To quantify the different levels of particle mobility underlying the two collective regimes, we consider the uncoupled dynamics ($\sigma=0$) in the absence of repulsive interactions ($\vec{F}=\vec{0}$). As shown in Fig.~\ref{fig:msd}, turbulent-chaotic heading dynamics leads to chaotic diffusion with a short-time diffusion coefficient that decays asymptotically as $D_{\mathrm{short}}\sim1/T$. Consequently, particle motion over a delay length becomes increasingly weak for large $T$, limiting neighbor reshuffling and leaving the interaction network sparse. In contrast, laminar-chaotic heading dynamics leads to run-and-tumble motion (cf.~\onlinecite{santra_run-and-tumble_2020}) with a diffusion coefficient that approaches a finite value as $T\to\infty$. The long nearly constant orientation plateaus characteristic of laminar chaos therefore produce persistent particle motion and efficient neighbor reshuffling, providing the rapid network mixing that underlies the self-sustained intermittent flocking mechanism described in the main text.

\section{Robustness of the main result}
\label{app:robustness}

\begin{figure}[h]
    \centering
    \includegraphics[width=1\linewidth]{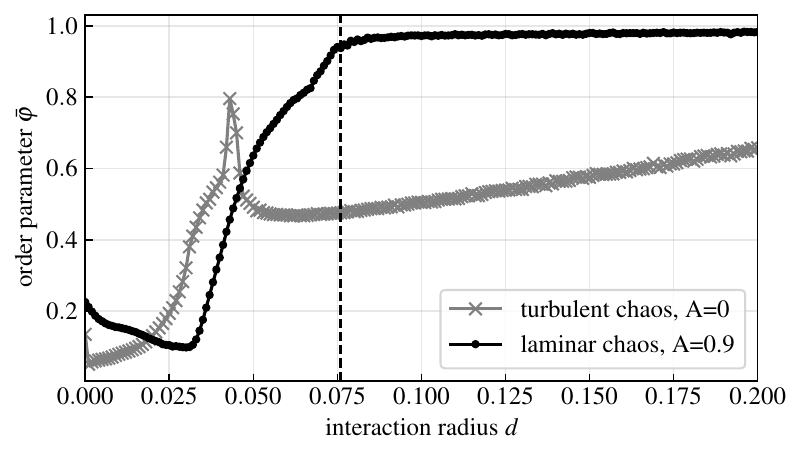}
    \caption{Robustness of the main result with respect to parameter changes. Time-averaged order parameter $\bar{\varphi}$ as a function of the interaction radius $d$ for turbulent and laminar chaos, where all other parameters are fixed at $\tau_0=1$, $N=1000$, $v=10$, $T=10^4$, $\sigma=1$, $\kappa=1$, and $d_F=0.038$. The dashed vertical line corresponds to the value $d=0.076$ used in the main text.}
    \label{fig:robustness}
\end{figure}

As demonstrated in Fig.~\ref{fig:robustness}, the qualitative difference between laminar and turbulent chaos does not rely on a finely tuned choice of the interaction radius. Varying $d$ while keeping the remaining parameters fixed preserves a broad separation between the two regimes: laminar chaos leads to nearly complete polar order over an extended range of interaction radii, whereas turbulent chaos remains substantially less ordered.

\bibliography{references}

\begin{thebibliography}{59}%
\makeatletter
\providecommand \@ifxundefined [1]{%
 \@ifx{#1\undefined}
}%
\providecommand \@ifnum [1]{%
 \ifnum #1\expandafter \@firstoftwo
 \else \expandafter \@secondoftwo
 \fi
}%
\providecommand \@ifx [1]{%
 \ifx #1\expandafter \@firstoftwo
 \else \expandafter \@secondoftwo
 \fi
}%
\providecommand \natexlab [1]{#1}%
\providecommand \enquote  [1]{``#1''}%
\providecommand \bibnamefont  [1]{#1}%
\providecommand \bibfnamefont [1]{#1}%
\providecommand \citenamefont [1]{#1}%
\providecommand \href@noop [0]{\@secondoftwo}%
\providecommand \href [0]{\begingroup \@sanitize@url \@href}%
\providecommand \@href[1]{\@@startlink{#1}\@@href}%
\providecommand \@@href[1]{\endgroup#1\@@endlink}%
\providecommand \@sanitize@url [0]{\catcode `\\12\catcode `\$12\catcode
  `\&12\catcode `\#12\catcode `\^12\catcode `\_12\catcode `\%12\relax}%
\providecommand \@@startlink[1]{}%
\providecommand \@@endlink[0]{}%
\providecommand \url  [0]{\begingroup\@sanitize@url \@url }%
\providecommand \@url [1]{\endgroup\@href {#1}{\urlprefix }}%
\providecommand \urlprefix  [0]{URL }%
\providecommand \Eprint [0]{\href }%
\providecommand \doibase [0]{http://dx.doi.org/}%
\providecommand \selectlanguage [0]{\@gobble}%
\providecommand \bibinfo  [0]{\@secondoftwo}%
\providecommand \bibfield  [0]{\@secondoftwo}%
\providecommand \translation [1]{[#1]}%
\providecommand \BibitemOpen [0]{}%
\providecommand \bibitemStop [0]{}%
\providecommand \bibitemNoStop [0]{.\EOS\space}%
\providecommand \EOS [0]{\spacefactor3000\relax}%
\providecommand \BibitemShut  [1]{\csname bibitem#1\endcsname}%
\let\auto@bib@innerbib\@empty
\bibitem [{\citenamefont {Cates}\ and\ \citenamefont
  {Tailleur}(2015)}]{Cates2015}%
  \BibitemOpen
  \bibfield  {author} {\bibinfo {author} {\bibfnamefont {M.~E.}\ \bibnamefont
  {Cates}}\ and\ \bibinfo {author} {\bibfnamefont {J.}~\bibnamefont
  {Tailleur}},\ }\bibfield  {title} {\enquote {\bibinfo {title}
  {Motility-induced phase separation},}\ }\href {\doibase
  10.1146/annurev-conmatphys-031214-014710} {\bibfield  {journal} {\bibinfo
  {journal} {Annual Review of Condensed Matter Physics}\ }\textbf {\bibinfo
  {volume} {6}},\ \bibinfo {pages} {219--244} (\bibinfo {year}
  {2015})}\BibitemShut {NoStop}%
\bibitem [{\citenamefont {Alert}, \citenamefont {Casademunt},\ and\
  \citenamefont {Joanny}(2022)}]{Alert2022}%
  \BibitemOpen
  \bibfield  {author} {\bibinfo {author} {\bibfnamefont {R.}~\bibnamefont
  {Alert}}, \bibinfo {author} {\bibfnamefont {J.}~\bibnamefont {Casademunt}}, \
  and\ \bibinfo {author} {\bibfnamefont {J.-F.}\ \bibnamefont {Joanny}},\
  }\bibfield  {title} {\enquote {\bibinfo {title} {Active turbulence},}\ }\href
  {\doibase 10.1146/annurev-conmatphys-082321-035957} {\bibfield  {journal}
  {\bibinfo  {journal} {Annual Review of Condensed Matter Physics}\ }\textbf
  {\bibinfo {volume} {13}},\ \bibinfo {pages} {143--170} (\bibinfo {year}
  {2022})}\BibitemShut {NoStop}%
\bibitem [{\citenamefont {Vicsek}\ \emph {et~al.}(1995)\citenamefont {Vicsek},
  \citenamefont {Czir\'ok}, \citenamefont {Ben-Jacob}, \citenamefont {Cohen},\
  and\ \citenamefont {Shochet}}]{Vicsek1995Phase}%
  \BibitemOpen
  \bibfield  {author} {\bibinfo {author} {\bibfnamefont {T.}~\bibnamefont
  {Vicsek}}, \bibinfo {author} {\bibfnamefont {A.}~\bibnamefont {Czir\'ok}},
  \bibinfo {author} {\bibfnamefont {E.}~\bibnamefont {Ben-Jacob}}, \bibinfo
  {author} {\bibfnamefont {I.}~\bibnamefont {Cohen}}, \ and\ \bibinfo {author}
  {\bibfnamefont {O.}~\bibnamefont {Shochet}},\ }\bibfield  {title} {\enquote
  {\bibinfo {title} {Novel type of phase transition in a system of self-driven
  particles},}\ }\href {\doibase 10.1103/PhysRevLett.75.1226} {\bibfield
  {journal} {\bibinfo  {journal} {Phys. Rev. Lett.}\ }\textbf {\bibinfo
  {volume} {75}},\ \bibinfo {pages} {1226--1229} (\bibinfo {year}
  {1995})}\BibitemShut {NoStop}%
\bibitem [{\citenamefont {Vicsek}\ and\ \citenamefont
  {Zafeiris}(2012)}]{Vicsek2012Collective}%
  \BibitemOpen
  \bibfield  {author} {\bibinfo {author} {\bibfnamefont {T.}~\bibnamefont
  {Vicsek}}\ and\ \bibinfo {author} {\bibfnamefont {A.}~\bibnamefont
  {Zafeiris}},\ }\bibfield  {title} {\enquote {\bibinfo {title} {Collective
  motion},}\ }\href {\doibase 10.1016/j.physrep.2012.03.004} {\bibfield
  {journal} {\bibinfo  {journal} {Physics Reports}\ }\textbf {\bibinfo {volume}
  {517}},\ \bibinfo {pages} {71--140} (\bibinfo {year} {2012})}\BibitemShut
  {NoStop}%
\bibitem [{\citenamefont {Frasca}\ \emph {et~al.}(2008)\citenamefont {Frasca},
  \citenamefont {Buscarino}, \citenamefont {Rizzo}, \citenamefont {Fortuna},\
  and\ \citenamefont {Boccaletti}}]{frasca_synchronization_2008}%
  \BibitemOpen
  \bibfield  {author} {\bibinfo {author} {\bibfnamefont {M.}~\bibnamefont
  {Frasca}}, \bibinfo {author} {\bibfnamefont {A.}~\bibnamefont {Buscarino}},
  \bibinfo {author} {\bibfnamefont {A.}~\bibnamefont {Rizzo}}, \bibinfo
  {author} {\bibfnamefont {L.}~\bibnamefont {Fortuna}}, \ and\ \bibinfo
  {author} {\bibfnamefont {S.}~\bibnamefont {Boccaletti}},\ }\bibfield  {title}
  {\enquote {\bibinfo {title} {Synchronization of {Moving} {Chaotic}
  {Agents}},}\ }\href {\doibase 10.1103/PhysRevLett.100.044102} {\bibfield
  {journal} {\bibinfo  {journal} {Phys. Rev. Lett.}\ }\textbf {\bibinfo
  {volume} {100}},\ \bibinfo {pages} {044102} (\bibinfo {year}
  {2008})}\BibitemShut {NoStop}%
\bibitem [{\citenamefont {Valani}\ and\ \citenamefont
  {Paganin}(2023)}]{Valaniattractormatter2023}%
  \BibitemOpen
  \bibfield  {author} {\bibinfo {author} {\bibfnamefont {R.~N.}\ \bibnamefont
  {Valani}}\ and\ \bibinfo {author} {\bibfnamefont {D.~M.}\ \bibnamefont
  {Paganin}},\ }\bibfield  {title} {\enquote {\bibinfo {title}
  {{Attractor-driven matter}},}\ }\href@noop {} {\bibfield  {journal} {\bibinfo
   {journal} {Chaos}\ }\textbf {\bibinfo {volume} {33}} (\bibinfo {year}
  {2023})},\ \bibinfo {note} {023125}\BibitemShut {NoStop}%
\bibitem [{\citenamefont {Sar}\ \emph {et~al.}(2026)\citenamefont {Sar},
  \citenamefont {O’Keeffe}, \citenamefont {Lizárraga}, \citenamefont {{de
  Aguiar}}, \citenamefont {Bettstetter},\ and\ \citenamefont
  {Ghosh}}]{SAR20261}%
  \BibitemOpen
  \bibfield  {author} {\bibinfo {author} {\bibfnamefont {G.~K.}\ \bibnamefont
  {Sar}}, \bibinfo {author} {\bibfnamefont {K.}~\bibnamefont {O’Keeffe}},
  \bibinfo {author} {\bibfnamefont {J.~U.}\ \bibnamefont {Lizárraga}},
  \bibinfo {author} {\bibfnamefont {M.~A.}\ \bibnamefont {{de Aguiar}}},
  \bibinfo {author} {\bibfnamefont {C.}~\bibnamefont {Bettstetter}}, \ and\
  \bibinfo {author} {\bibfnamefont {D.}~\bibnamefont {Ghosh}},\ }\bibfield
  {title} {\enquote {\bibinfo {title} {Interplay of sync and swarm: Theory and
  application of swarmalators},}\ }\href {\doibase
  https://doi.org/10.1016/j.physrep.2026.01.002} {\bibfield  {journal}
  {\bibinfo  {journal} {Physics Reports}\ }\textbf {\bibinfo {volume} {1167}},\
  \bibinfo {pages} {1--52} (\bibinfo {year} {2026})}\BibitemShut {NoStop}%
\bibitem [{\citenamefont {O'Keeffe}, \citenamefont {Hong},\ and\ \citenamefont
  {Strogatz}(2017)}]{OKeeffe2017}%
  \BibitemOpen
  \bibfield  {author} {\bibinfo {author} {\bibfnamefont {K.~P.}\ \bibnamefont
  {O'Keeffe}}, \bibinfo {author} {\bibfnamefont {H.}~\bibnamefont {Hong}}, \
  and\ \bibinfo {author} {\bibfnamefont {S.~H.}\ \bibnamefont {Strogatz}},\
  }\bibfield  {title} {\enquote {\bibinfo {title} {Oscillators that sync and
  swarm},}\ }\href {\doibase 10.1038/s41467-017-01190-3} {\bibfield  {journal}
  {\bibinfo  {journal} {Nature Communications}\ }\textbf {\bibinfo {volume}
  {8}},\ \bibinfo {pages} {1504} (\bibinfo {year} {2017})}\BibitemShut
  {NoStop}%
\bibitem [{\citenamefont {Gei\ss{}}, \citenamefont {Kroy},\ and\ \citenamefont
  {Holubec}(2022)}]{PhysRevE.106.054612}%
  \BibitemOpen
  \bibfield  {author} {\bibinfo {author} {\bibfnamefont {D.}~\bibnamefont
  {Gei\ss{}}}, \bibinfo {author} {\bibfnamefont {K.}~\bibnamefont {Kroy}}, \
  and\ \bibinfo {author} {\bibfnamefont {V.}~\bibnamefont {Holubec}},\
  }\bibfield  {title} {\enquote {\bibinfo {title} {Signal propagation and
  linear response in the delay {V}icsek model},}\ }\href {\doibase
  10.1103/PhysRevE.106.054612} {\bibfield  {journal} {\bibinfo  {journal}
  {Phys. Rev. E}\ }\textbf {\bibinfo {volume} {106}},\ \bibinfo {pages}
  {054612} (\bibinfo {year} {2022})}\BibitemShut {NoStop}%
\bibitem [{\citenamefont {Holubec}\ \emph {et~al.}(2021)\citenamefont
  {Holubec}, \citenamefont {Geiss}, \citenamefont {Loos}, \citenamefont
  {Kroy},\ and\ \citenamefont {Cichos}}]{PhysRevLett.127.258001}%
  \BibitemOpen
  \bibfield  {author} {\bibinfo {author} {\bibfnamefont {V.}~\bibnamefont
  {Holubec}}, \bibinfo {author} {\bibfnamefont {D.}~\bibnamefont {Geiss}},
  \bibinfo {author} {\bibfnamefont {S.~A.~M.}\ \bibnamefont {Loos}}, \bibinfo
  {author} {\bibfnamefont {K.}~\bibnamefont {Kroy}}, \ and\ \bibinfo {author}
  {\bibfnamefont {F.}~\bibnamefont {Cichos}},\ }\bibfield  {title} {\enquote
  {\bibinfo {title} {Finite-size scaling at the edge of disorder in a
  time-delay {V}icsek model},}\ }\href {\doibase
  10.1103/PhysRevLett.127.258001} {\bibfield  {journal} {\bibinfo  {journal}
  {Phys. Rev. Lett.}\ }\textbf {\bibinfo {volume} {127}},\ \bibinfo {pages}
  {258001} (\bibinfo {year} {2021})}\BibitemShut {NoStop}%
\bibitem [{\citenamefont {Horton}\ and\ \citenamefont
  {Holubec}(2025)}]{Horton_2025}%
  \BibitemOpen
  \bibfield  {author} {\bibinfo {author} {\bibfnamefont {R.}~\bibnamefont
  {Horton}}\ and\ \bibinfo {author} {\bibfnamefont {V.}~\bibnamefont
  {Holubec}},\ }\bibfield  {title} {\enquote {\bibinfo {title} {Order-disorder
  transition and phase separation in delay {V}icsek model},}\ }\href {\doibase
  10.1088/1367-2630/ae02be} {\bibfield  {journal} {\bibinfo  {journal} {New
  Journal of Physics}\ }\textbf {\bibinfo {volume} {27}},\ \bibinfo {pages}
  {094402} (\bibinfo {year} {2025})}\BibitemShut {NoStop}%
\bibitem [{\citenamefont {Pakpour}\ and\ \citenamefont
  {Vicsek}(2024)}]{PAKPOUR2024129453}%
  \BibitemOpen
  \bibfield  {author} {\bibinfo {author} {\bibfnamefont {F.}~\bibnamefont
  {Pakpour}}\ and\ \bibinfo {author} {\bibfnamefont {T.}~\bibnamefont
  {Vicsek}},\ }\bibfield  {title} {\enquote {\bibinfo {title} {Delay-induced
  phase transitions in active matter},}\ }\href {\doibase
  https://doi.org/10.1016/j.physa.2023.129453} {\bibfield  {journal} {\bibinfo
  {journal} {Physica A: Statistical Mechanics and its Applications}\ }\textbf
  {\bibinfo {volume} {634}},\ \bibinfo {pages} {129453} (\bibinfo {year}
  {2024})}\BibitemShut {NoStop}%
\bibitem [{\citenamefont {Sun}, \citenamefont {Lin},\ and\ \citenamefont
  {Erban}(2014)}]{PhysRevE.90.062708}%
  \BibitemOpen
  \bibfield  {author} {\bibinfo {author} {\bibfnamefont {Y.}~\bibnamefont
  {Sun}}, \bibinfo {author} {\bibfnamefont {W.}~\bibnamefont {Lin}}, \ and\
  \bibinfo {author} {\bibfnamefont {R.}~\bibnamefont {Erban}},\ }\bibfield
  {title} {\enquote {\bibinfo {title} {Time delay can facilitate coherence in
  self-driven interacting-particle systems},}\ }\href {\doibase
  10.1103/PhysRevE.90.062708} {\bibfield  {journal} {\bibinfo  {journal} {Phys.
  Rev. E}\ }\textbf {\bibinfo {volume} {90}},\ \bibinfo {pages} {062708}
  (\bibinfo {year} {2014})}\BibitemShut {NoStop}%
\bibitem [{\citenamefont {Costanzo}, \citenamefont {van Haeringen},\ and\
  \citenamefont {Hemelrijk}(2022)}]{Costanzo_2022}%
  \BibitemOpen
  \bibfield  {author} {\bibinfo {author} {\bibfnamefont {A.}~\bibnamefont
  {Costanzo}}, \bibinfo {author} {\bibfnamefont {E.}~\bibnamefont {van
  Haeringen}}, \ and\ \bibinfo {author} {\bibfnamefont {C.~K.}\ \bibnamefont
  {Hemelrijk}},\ }\bibfield  {title} {\enquote {\bibinfo {title} {Effect of
  time-delayed interactions on milling: A minimal model},}\ }\href {\doibase
  10.1209/0295-5075/ac5ed1} {\bibfield  {journal} {\bibinfo  {journal}
  {Europhysics Letters}\ }\textbf {\bibinfo {volume} {138}},\ \bibinfo {pages}
  {22002} (\bibinfo {year} {2022})}\BibitemShut {NoStop}%
\bibitem [{\citenamefont {Szwaykowska}\ \emph {et~al.}(2016)\citenamefont
  {Szwaykowska}, \citenamefont {Schwartz}, \citenamefont {Mier-y Teran~Romero},
  \citenamefont {Heckman}, \citenamefont {Mox},\ and\ \citenamefont
  {Hsieh}}]{PhysRevE.93.032307}%
  \BibitemOpen
  \bibfield  {author} {\bibinfo {author} {\bibfnamefont {K.}~\bibnamefont
  {Szwaykowska}}, \bibinfo {author} {\bibfnamefont {I.~B.}\ \bibnamefont
  {Schwartz}}, \bibinfo {author} {\bibfnamefont {L.}~\bibnamefont {Mier-y
  Teran~Romero}}, \bibinfo {author} {\bibfnamefont {C.~R.}\ \bibnamefont
  {Heckman}}, \bibinfo {author} {\bibfnamefont {D.}~\bibnamefont {Mox}}, \ and\
  \bibinfo {author} {\bibfnamefont {M.~A.}\ \bibnamefont {Hsieh}},\ }\bibfield
  {title} {\enquote {\bibinfo {title} {Collective motion patterns of swarms
  with delay coupling: Theory and experiment},}\ }\href {\doibase
  10.1103/PhysRevE.93.032307} {\bibfield  {journal} {\bibinfo  {journal} {Phys.
  Rev. E}\ }\textbf {\bibinfo {volume} {93}},\ \bibinfo {pages} {032307}
  (\bibinfo {year} {2016})}\BibitemShut {NoStop}%
\bibitem [{\citenamefont {Chen}, \citenamefont {L{\"u}},\ and\ \citenamefont
  {Yu}(2011)}]{Chen2011}%
  \BibitemOpen
  \bibfield  {author} {\bibinfo {author} {\bibfnamefont {Y.}~\bibnamefont
  {Chen}}, \bibinfo {author} {\bibfnamefont {J.}~\bibnamefont {L{\"u}}}, \ and\
  \bibinfo {author} {\bibfnamefont {X.}~\bibnamefont {Yu}},\ }\bibfield
  {title} {\enquote {\bibinfo {title} {Robust consensus of multi-agent systems
  with time-varying delays in noisy environment},}\ }\href {\doibase
  10.1007/s11431-011-4477-y} {\bibfield  {journal} {\bibinfo  {journal}
  {Science China Technological Sciences}\ }\textbf {\bibinfo {volume} {54}},\
  \bibinfo {pages} {2014--2023} (\bibinfo {year} {2011})}\BibitemShut {NoStop}%
\bibitem [{\citenamefont {Chen}, \citenamefont {Lü},\ and\ \citenamefont
  {Lin}(2009)}]{5400490}%
  \BibitemOpen
  \bibfield  {author} {\bibinfo {author} {\bibfnamefont {Y.}~\bibnamefont
  {Chen}}, \bibinfo {author} {\bibfnamefont {J.}~\bibnamefont {Lü}}, \ and\
  \bibinfo {author} {\bibfnamefont {Z.}~\bibnamefont {Lin}},\ }\bibfield
  {title} {\enquote {\bibinfo {title} {Consensus of discrete-time multi-agent
  systems with nonlinear local rules and time-varying delays},}\ }in\ \href
  {\doibase 10.1109/CDC.2009.5400490} {\emph {\bibinfo {booktitle} {Proceedings
  of the 48th IEEE Conference on Decision and Control (CDC) held jointly with
  2009 28th Chinese Control Conference}}}\ (\bibinfo {year} {2009})\ pp.\
  \bibinfo {pages} {7018--7023}\BibitemShut {NoStop}%
\bibitem [{\citenamefont {Li}\ \emph {et~al.}(2026{\natexlab{a}})\citenamefont
  {Li}, \citenamefont {Shi}, \citenamefont {Shi}, \citenamefont {Lin},\ and\
  \citenamefont {Qin}}]{11489308}%
  \BibitemOpen
  \bibfield  {author} {\bibinfo {author} {\bibfnamefont {W.}~\bibnamefont
  {Li}}, \bibinfo {author} {\bibfnamefont {M.}~\bibnamefont {Shi}}, \bibinfo
  {author} {\bibfnamefont {L.}~\bibnamefont {Shi}}, \bibinfo {author}
  {\bibfnamefont {B.}~\bibnamefont {Lin}}, \ and\ \bibinfo {author}
  {\bibfnamefont {K.}~\bibnamefont {Qin}},\ }\bibfield  {title} {\enquote
  {\bibinfo {title} {Collective flocking of networked swarms with fading
  channels and time-varying delays},}\ }\href {\doibase
  10.1109/TGCN.2026.3686027} {\bibfield  {journal} {\bibinfo  {journal} {IEEE
  Transactions on Green Communications and Networking}\ }\textbf {\bibinfo
  {volume} {10}},\ \bibinfo {pages} {2766--2777} (\bibinfo {year}
  {2026}{\natexlab{a}})}\BibitemShut {NoStop}%
\bibitem [{\citenamefont {Xie}, \citenamefont {Zhong},\ and\ \citenamefont
  {Cui}(2025)}]{10.1145/3760269.3760339}%
  \BibitemOpen
  \bibfield  {author} {\bibinfo {author} {\bibfnamefont {J.}~\bibnamefont
  {Xie}}, \bibinfo {author} {\bibfnamefont {Y.}~\bibnamefont {Zhong}}, \ and\
  \bibinfo {author} {\bibfnamefont {Y.}~\bibnamefont {Cui}},\ }\bibfield
  {title} {\enquote {\bibinfo {title} {Advances and future directions in
  flocking control models for multi-agent systems},}\ }in\ \href {\doibase
  10.1145/3760269.3760339} {\emph {\bibinfo {booktitle} {Proceedings of the
  2025 5th International Conference on Automation Control, Algorithm and
  Intelligent Bionics}}},\ \bibinfo {series and number} {ACAIB '25}\ (\bibinfo
  {publisher} {Association for Computing Machinery},\ \bibinfo {address} {New
  York, NY, USA},\ \bibinfo {year} {2025})\ p.\ \bibinfo {pages}
  {437–443}\BibitemShut {NoStop}%
\bibitem [{\citenamefont {Li}\ \emph {et~al.}(2026{\natexlab{b}})\citenamefont
  {Li}, \citenamefont {Phan}, \citenamefont {Di~Carlo}, \citenamefont {Wang},
  \citenamefont {Do}, \citenamefont {Mikhail}, \citenamefont {Austin},\ and\
  \citenamefont {Liu}}]{li_informational_2026}%
  \BibitemOpen
  \bibfield  {author} {\bibinfo {author} {\bibfnamefont {S.}~\bibnamefont
  {Li}}, \bibinfo {author} {\bibfnamefont {T.~V.}\ \bibnamefont {Phan}},
  \bibinfo {author} {\bibfnamefont {L.}~\bibnamefont {Di~Carlo}}, \bibinfo
  {author} {\bibfnamefont {G.}~\bibnamefont {Wang}}, \bibinfo {author}
  {\bibfnamefont {V.~H.}\ \bibnamefont {Do}}, \bibinfo {author} {\bibfnamefont
  {E.}~\bibnamefont {Mikhail}}, \bibinfo {author} {\bibfnamefont {R.~H.}\
  \bibnamefont {Austin}}, \ and\ \bibinfo {author} {\bibfnamefont
  {L.}~\bibnamefont {Liu}},\ }\bibfield  {title} {\enquote {\bibinfo {title}
  {Informational {Memory} {Shapes} {Collective} {Behavior} in {Intelligent}
  {Swarms}},}\ }\href {\doibase 10.1103/rt97-ncmf} {\bibfield  {journal}
  {\bibinfo  {journal} {Phys. Rev. Lett.}\ }\textbf {\bibinfo {volume} {136}},\
  \bibinfo {pages} {138302} (\bibinfo {year} {2026}{\natexlab{b}})}\BibitemShut
  {NoStop}%
\bibitem [{\citenamefont {Müller}, \citenamefont {Otto},\ and\ \citenamefont
  {Radons}(2018)}]{muller_laminar_2018}%
  \BibitemOpen
  \bibfield  {author} {\bibinfo {author} {\bibfnamefont {D.}~\bibnamefont
  {Müller}}, \bibinfo {author} {\bibfnamefont {A.}~\bibnamefont {Otto}}, \
  and\ \bibinfo {author} {\bibfnamefont {G.}~\bibnamefont {Radons}},\
  }\bibfield  {title} {\enquote {\bibinfo {title} {Laminar chaos},}\ }\href
  {\doibase 10.1103/PhysRevLett.120.084102} {\bibfield  {journal} {\bibinfo
  {journal} {Phys. Rev. Lett.}\ }\textbf {\bibinfo {volume} {120}},\ \bibinfo
  {pages} {084102} (\bibinfo {year} {2018})}\BibitemShut {NoStop}%
\bibitem [{\citenamefont {Fujiwara}, \citenamefont {Kurths},\ and\
  \citenamefont {Díaz-Guilera}(2016)}]{fujiwara_synchronization_2016}%
  \BibitemOpen
  \bibfield  {author} {\bibinfo {author} {\bibfnamefont {N.}~\bibnamefont
  {Fujiwara}}, \bibinfo {author} {\bibfnamefont {J.}~\bibnamefont {Kurths}}, \
  and\ \bibinfo {author} {\bibfnamefont {A.}~\bibnamefont {Díaz-Guilera}},\
  }\bibfield  {title} {\enquote {\bibinfo {title} {Synchronization of mobile
  chaotic oscillator networks},}\ }\href {\doibase 10.1063/1.4962129}
  {\bibfield  {journal} {\bibinfo  {journal} {Chaos}\ }\textbf {\bibinfo
  {volume} {26}},\ \bibinfo {pages} {094824} (\bibinfo {year}
  {2016})}\BibitemShut {NoStop}%
\bibitem [{\citenamefont {Glass}, \citenamefont {Jin},\ and\ \citenamefont
  {Riedel-Kruse}(2021)}]{glass_nonlinear_2021}%
  \BibitemOpen
  \bibfield  {author} {\bibinfo {author} {\bibfnamefont {D.~S.}\ \bibnamefont
  {Glass}}, \bibinfo {author} {\bibfnamefont {X.}~\bibnamefont {Jin}}, \ and\
  \bibinfo {author} {\bibfnamefont {I.~H.}\ \bibnamefont {Riedel-Kruse}},\
  }\bibfield  {title} {\enquote {\bibinfo {title} {Nonlinear delay differential
  equations and their application to modeling biological network motifs},}\
  }\href {\doibase 10.1038/s41467-021-21700-8} {\bibfield  {journal} {\bibinfo
  {journal} {Nat. Commun.}\ }\textbf {\bibinfo {volume} {12}},\ \bibinfo
  {pages} {1788} (\bibinfo {year} {2021})}\BibitemShut {NoStop}%
\bibitem [{\citenamefont {Mackey}\ and\ \citenamefont
  {Glass}(1977)}]{mackey_oscillation_1977}%
  \BibitemOpen
  \bibfield  {author} {\bibinfo {author} {\bibfnamefont {M.~C.}\ \bibnamefont
  {Mackey}}\ and\ \bibinfo {author} {\bibfnamefont {L.}~\bibnamefont {Glass}},\
  }\bibfield  {title} {\enquote {\bibinfo {title} {Oscillation and chaos in
  physiological control systems},}\ }\href {\doibase 10.1126/science.267326}
  {\bibfield  {journal} {\bibinfo  {journal} {Science}\ }\textbf {\bibinfo
  {volume} {197}},\ \bibinfo {pages} {287--289} (\bibinfo {year}
  {1977})}\BibitemShut {NoStop}%
\bibitem [{\citenamefont {Ikeda}(1979)}]{ikeda_multiple-valued_1979}%
  \BibitemOpen
  \bibfield  {author} {\bibinfo {author} {\bibfnamefont {K.}~\bibnamefont
  {Ikeda}},\ }\bibfield  {title} {\enquote {\bibinfo {title} {Multiple-valued
  stationary state and its instability of the transmitted light by a ring
  cavity system},}\ }\href {\doibase 10.1016/0030-4018(79)90090-7} {\bibfield
  {journal} {\bibinfo  {journal} {Opt. Commun.}\ }\textbf {\bibinfo {volume}
  {30}},\ \bibinfo {pages} {257--261} (\bibinfo {year} {1979})}\BibitemShut
  {NoStop}%
\bibitem [{\citenamefont {Ikeda}, \citenamefont {Daido},\ and\ \citenamefont
  {Akimoto}(1980)}]{ikeda_optical_1980}%
  \BibitemOpen
  \bibfield  {author} {\bibinfo {author} {\bibfnamefont {K.}~\bibnamefont
  {Ikeda}}, \bibinfo {author} {\bibfnamefont {H.}~\bibnamefont {Daido}}, \ and\
  \bibinfo {author} {\bibfnamefont {O.}~\bibnamefont {Akimoto}},\ }\bibfield
  {title} {\enquote {\bibinfo {title} {Optical {Turbulence}: {Chaotic}
  {Behavior} of {Transmitted} {Light} from a {Ring} {Cavity}},}\ }\href
  {\doibase 10.1103/PhysRevLett.45.709} {\bibfield  {journal} {\bibinfo
  {journal} {Phys. Rev. Lett.}\ }\textbf {\bibinfo {volume} {45}},\ \bibinfo
  {pages} {709--712} (\bibinfo {year} {1980})}\BibitemShut {NoStop}%
\bibitem [{\citenamefont {Hart}\ \emph
  {et~al.}(2019{\natexlab{a}})\citenamefont {Hart}, \citenamefont {Roy},
  \citenamefont {Müller-Bender}, \citenamefont {Otto},\ and\ \citenamefont
  {Radons}}]{hart_laminar_2019}%
  \BibitemOpen
  \bibfield  {author} {\bibinfo {author} {\bibfnamefont {J.~D.}\ \bibnamefont
  {Hart}}, \bibinfo {author} {\bibfnamefont {R.}~\bibnamefont {Roy}}, \bibinfo
  {author} {\bibfnamefont {D.}~\bibnamefont {Müller-Bender}}, \bibinfo
  {author} {\bibfnamefont {A.}~\bibnamefont {Otto}}, \ and\ \bibinfo {author}
  {\bibfnamefont {G.}~\bibnamefont {Radons}},\ }\bibfield  {title} {\enquote
  {\bibinfo {title} {Laminar chaos in experiments: {Nonlinear} systems with
  time-varying delays and noise},}\ }\href {\doibase
  10.1103/PhysRevLett.123.154101} {\bibfield  {journal} {\bibinfo  {journal}
  {Phys. Rev. Lett.}\ }\textbf {\bibinfo {volume} {123}},\ \bibinfo {pages}
  {154101} (\bibinfo {year} {2019}{\natexlab{a}})}\BibitemShut {NoStop}%
\bibitem [{\citenamefont {Larger}(2013)}]{larger_complexity_2013}%
  \BibitemOpen
  \bibfield  {author} {\bibinfo {author} {\bibfnamefont {L.}~\bibnamefont
  {Larger}},\ }\bibfield  {title} {\enquote {\bibinfo {title} {Complexity in
  electro-optic delay dynamics: modelling, design and applications},}\ }\href
  {\doibase 10.1098/rsta.2012.0464} {\bibfield  {journal} {\bibinfo  {journal}
  {Phil. Trans. R. Soc. A}\ }\textbf {\bibinfo {volume} {371}},\ \bibinfo
  {pages} {20120464} (\bibinfo {year} {2013})}\BibitemShut {NoStop}%
\bibitem [{\citenamefont {Chembo}\ \emph {et~al.}(2019)\citenamefont {Chembo},
  \citenamefont {Brunner}, \citenamefont {Jacquot},\ and\ \citenamefont
  {Larger}}]{chembo_optoelectronic_2019}%
  \BibitemOpen
  \bibfield  {author} {\bibinfo {author} {\bibfnamefont {Y.~K.}\ \bibnamefont
  {Chembo}}, \bibinfo {author} {\bibfnamefont {D.}~\bibnamefont {Brunner}},
  \bibinfo {author} {\bibfnamefont {M.}~\bibnamefont {Jacquot}}, \ and\
  \bibinfo {author} {\bibfnamefont {L.}~\bibnamefont {Larger}},\ }\bibfield
  {title} {\enquote {\bibinfo {title} {Optoelectronic oscillators with
  time-delayed feedback},}\ }\href {\doibase 10.1103/RevModPhys.91.035006}
  {\bibfield  {journal} {\bibinfo  {journal} {Rev. Mod. Phys.}\ }\textbf
  {\bibinfo {volume} {91}},\ \bibinfo {pages} {035006} (\bibinfo {year}
  {2019})}\BibitemShut {NoStop}%
\bibitem [{\citenamefont {Sysoev}\ \emph {et~al.}(2016)\citenamefont {Sysoev},
  \citenamefont {Ponomarenko}, \citenamefont {Kulminskiy},\ and\ \citenamefont
  {Prokhorov}}]{sysoev_recovery_2016}%
  \BibitemOpen
  \bibfield  {author} {\bibinfo {author} {\bibfnamefont {I.~V.}\ \bibnamefont
  {Sysoev}}, \bibinfo {author} {\bibfnamefont {V.~I.}\ \bibnamefont
  {Ponomarenko}}, \bibinfo {author} {\bibfnamefont {D.~D.}\ \bibnamefont
  {Kulminskiy}}, \ and\ \bibinfo {author} {\bibfnamefont {M.~D.}\ \bibnamefont
  {Prokhorov}},\ }\bibfield  {title} {\enquote {\bibinfo {title} {Recovery of
  couplings and parameters of elements in networks of time-delay systems from
  time series},}\ }\href {\doibase 10.1103/PhysRevE.94.052207} {\bibfield
  {journal} {\bibinfo  {journal} {Phys. Rev. E}\ }\textbf {\bibinfo {volume}
  {94}},\ \bibinfo {pages} {052207} (\bibinfo {year} {2016})}\BibitemShut
  {NoStop}%
\bibitem [{\citenamefont {Ponomarenko}, \citenamefont {Kulminskiy},\ and\
  \citenamefont {Prokhorov}(2017)}]{ponomarenko_chimeralike_2017}%
  \BibitemOpen
  \bibfield  {author} {\bibinfo {author} {\bibfnamefont {V.~I.}\ \bibnamefont
  {Ponomarenko}}, \bibinfo {author} {\bibfnamefont {D.~D.}\ \bibnamefont
  {Kulminskiy}}, \ and\ \bibinfo {author} {\bibfnamefont {M.~D.}\ \bibnamefont
  {Prokhorov}},\ }\bibfield  {title} {\enquote {\bibinfo {title} {Chimeralike
  states in networks of bistable time-delayed feedback oscillators coupled via
  the mean field},}\ }\href {\doibase 10.1103/PhysRevE.96.022209} {\bibfield
  {journal} {\bibinfo  {journal} {Phys. Rev. E}\ }\textbf {\bibinfo {volume}
  {96}},\ \bibinfo {pages} {022209} (\bibinfo {year} {2017})}\BibitemShut
  {NoStop}%
\bibitem [{\citenamefont {Müller-Bender}\ and\ \citenamefont
  {Radons}(2023)}]{muller-bender_laminar_2023}%
  \BibitemOpen
  \bibfield  {author} {\bibinfo {author} {\bibfnamefont {D.}~\bibnamefont
  {Müller-Bender}}\ and\ \bibinfo {author} {\bibfnamefont {G.}~\bibnamefont
  {Radons}},\ }\bibfield  {title} {\enquote {\bibinfo {title} {Laminar chaos in
  systems with quasiperiodic delay},}\ }\href {\doibase
  10.1103/PhysRevE.107.014205} {\bibfield  {journal} {\bibinfo  {journal}
  {Phys. Rev. E}\ }\textbf {\bibinfo {volume} {107}},\ \bibinfo {pages}
  {014205} (\bibinfo {year} {2023})}\BibitemShut {NoStop}%
\bibitem [{\citenamefont {Müller-Bender}\ and\ \citenamefont
  {Valani}(2025)}]{muller-bender_laminar_2025}%
  \BibitemOpen
  \bibfield  {author} {\bibinfo {author} {\bibfnamefont {D.}~\bibnamefont
  {Müller-Bender}}\ and\ \bibinfo {author} {\bibfnamefont {R.~N.}\
  \bibnamefont {Valani}},\ }\bibfield  {title} {\enquote {\bibinfo {title}
  {Laminar chaos in systems with random and chaotically time-varying delay},}\
  }\href {\doibase 10.1103/cwjk-n45m} {\bibfield  {journal} {\bibinfo
  {journal} {Phys. Rev. E}\ }\textbf {\bibinfo {volume} {112}},\ \bibinfo
  {pages} {064203} (\bibinfo {year} {2025})}\BibitemShut {NoStop}%
\bibitem [{\citenamefont {Albers}\ \emph {et~al.}(2022)\citenamefont {Albers},
  \citenamefont {Müller-Bender}, \citenamefont {Hille},\ and\ \citenamefont
  {Radons}}]{albers_chaotic_2022}%
  \BibitemOpen
  \bibfield  {author} {\bibinfo {author} {\bibfnamefont {T.}~\bibnamefont
  {Albers}}, \bibinfo {author} {\bibfnamefont {D.}~\bibnamefont
  {Müller-Bender}}, \bibinfo {author} {\bibfnamefont {L.}~\bibnamefont
  {Hille}}, \ and\ \bibinfo {author} {\bibfnamefont {G.}~\bibnamefont
  {Radons}},\ }\bibfield  {title} {\enquote {\bibinfo {title} {Chaotic
  {Diffusion} in {Delay} {Systems}: {Giant} {Enhancement} by {Time} {Lag}
  {Modulation}},}\ }\href {\doibase 10.1103/PhysRevLett.128.074101} {\bibfield
  {journal} {\bibinfo  {journal} {Phys. Rev. Lett.}\ }\textbf {\bibinfo
  {volume} {128}},\ \bibinfo {pages} {074101} (\bibinfo {year}
  {2022})}\BibitemShut {NoStop}%
\bibitem [{Note1()}]{Note1}%
  \BibitemOpen
  \bibinfo {note} {For the numerical results, Eqs.~\protect \eqref
  {eq:position_ode} and \protect \eqref {eq:heading_dde} were solved using the
  trapezoidal rule and the Lobatto IIIC method \cite {bellen_numerical_2003},
  respectively, where linear interpolation was used for the delayed term. We
  used a solver time step $\Delta t = (10\protect \,T)^{-1}$ and initialized
  the system with random constant initial history functions uniformly
  distributed in $[0,1]\times [0,1]$ for the position and $[0,2\pi ]$ for the
  heading. Before computing the observables, we let the transients relax for at
  least $1000$ delay periods. Drafting of the source code was assisted by
  ChatGPT 5.3-5.5, where the code was reviewed and tested
  manually.}\BibitemShut {Stop}%
\bibitem [{Note2()}]{Note2}%
  \BibitemOpen
  \bibinfo {note} {The time averages were computed from time series of length
  $t_\protect \mathrm {end}-t_0=1000$ after discarding an initial transient of
  duration $t_0=1000$. The time step used for the averaging was $10^{-2}$,
  while the numerical integration time step was $10^{-5}$.}\BibitemShut {Stop}%
\bibitem [{\citenamefont {Otto}, \citenamefont {Müller},\ and\ \citenamefont
  {Radons}(2017)}]{otto_universal_2017}%
  \BibitemOpen
  \bibfield  {author} {\bibinfo {author} {\bibfnamefont {A.}~\bibnamefont
  {Otto}}, \bibinfo {author} {\bibfnamefont {D.}~\bibnamefont {Müller}}, \
  and\ \bibinfo {author} {\bibfnamefont {G.}~\bibnamefont {Radons}},\
  }\bibfield  {title} {\enquote {\bibinfo {title} {Universal dichotomy for
  dynamical systems with variable delay},}\ }\href {\doibase
  10.1103/PhysRevLett.118.044104} {\bibfield  {journal} {\bibinfo  {journal}
  {Phys. Rev. Lett.}\ }\textbf {\bibinfo {volume} {118}},\ \bibinfo {pages}
  {044104} (\bibinfo {year} {2017})}\BibitemShut {NoStop}%
\bibitem [{\citenamefont {Katok}\ and\ \citenamefont
  {Hasselblatt}(1997)}]{katok_introduction_1997}%
  \BibitemOpen
  \bibfield  {author} {\bibinfo {author} {\bibfnamefont {A.}~\bibnamefont
  {Katok}}\ and\ \bibinfo {author} {\bibfnamefont {B.}~\bibnamefont
  {Hasselblatt}},\ }\href {\doibase 10.1017/CBO9780511809187} {\emph {\bibinfo
  {title} {Introduction to the modern theory of dynamical systems}}},\ \bibinfo
  {series} {Encyclopedia of Mathematics and its Applications}, Vol.~\bibinfo
  {volume} {54}\ (\bibinfo  {publisher} {Cambridge University Press},\ \bibinfo
  {address} {Cambridge},\ \bibinfo {year} {1997})\BibitemShut {NoStop}%
\bibitem [{\citenamefont {Arnold}(1961)}]{arnold_small_1961}%
  \BibitemOpen
  \bibfield  {author} {\bibinfo {author} {\bibfnamefont {V.~I.}\ \bibnamefont
  {Arnold}},\ }\bibfield  {title} {\enquote {\bibinfo {title} {Small
  denominators {I}: Mappings of the circle onto itself},}\ }\href
  {http://mi.mathnet.ru/izv3366} {\bibfield  {journal} {\bibinfo  {journal}
  {Izvest. Akad. Nauk SSSR Ser. Mat.}\ }\textbf {\bibinfo {volume} {25}},\
  \bibinfo {pages} {21--86} (\bibinfo {year} {1961})},\ \translation{Amer.
  Math. Soc. Transl. Ser. 2 \textbf{46}, 213 (1965)}\BibitemShut {NoStop}%
\bibitem [{\citenamefont {Arnold}(1964)}]{arnold_small_1961_erratum}%
  \BibitemOpen
  \bibfield  {author} {\bibinfo {author} {\bibfnamefont {V.~I.}\ \bibnamefont
  {Arnold}},\ }\href {http://mi.mathnet.ru/izv2961} {\bibfield  {journal}
  {\bibinfo  {journal} {Izvest. Akad. Nauk SSSR Ser. Mat.}\ }\textbf {\bibinfo
  {volume} {28}},\ \bibinfo {pages} {479--480} (\bibinfo {year} {1964})},\
  \bibinfo {note} {{Erratum}}\BibitemShut {NoStop}%
\bibitem [{\citenamefont {Ott}(2002)}]{ott_chaos_2002}%
  \BibitemOpen
  \bibfield  {author} {\bibinfo {author} {\bibfnamefont {E.}~\bibnamefont
  {Ott}},\ }\href {\doibase 10.1017/CBO9780511803260} {\emph {\bibinfo {title}
  {Chaos in dynamical systems}}}\ (\bibinfo  {publisher} {Cambridge University
  Press},\ \bibinfo {address} {Cambridge},\ \bibinfo {year} {2002})\BibitemShut
  {NoStop}%
\bibitem [{\citenamefont {Müller-Bender}, \citenamefont {Otto},\ and\
  \citenamefont {Radons}(2019)}]{muller-bender_resonant_2019}%
  \BibitemOpen
  \bibfield  {author} {\bibinfo {author} {\bibfnamefont {D.}~\bibnamefont
  {Müller-Bender}}, \bibinfo {author} {\bibfnamefont {A.}~\bibnamefont
  {Otto}}, \ and\ \bibinfo {author} {\bibfnamefont {G.}~\bibnamefont
  {Radons}},\ }\bibfield  {title} {\enquote {\bibinfo {title} {Resonant
  {Doppler} effect in systems with variable delay},}\ }\href {\doibase
  10.1098/rsta.2018.0119} {\bibfield  {journal} {\bibinfo  {journal} {Phil.
  Trans. R. Soc. A}\ }\textbf {\bibinfo {volume} {377}},\ \bibinfo {pages}
  {20180119} (\bibinfo {year} {2019})}\BibitemShut {NoStop}%
\bibitem [{\citenamefont {Bellman}\ and\ \citenamefont
  {Cooke}(1965)}]{bellman_computational_1965}%
  \BibitemOpen
  \bibfield  {author} {\bibinfo {author} {\bibfnamefont {R.}~\bibnamefont
  {Bellman}}\ and\ \bibinfo {author} {\bibfnamefont {K.~L.}\ \bibnamefont
  {Cooke}},\ }\bibfield  {title} {\enquote {\bibinfo {title} {On the
  computational solution of a class of functional differential equations},}\
  }\href {\doibase 10.1016/0022-247X(65)90017-X} {\bibfield  {journal}
  {\bibinfo  {journal} {J. Math. Anal. Appl.}\ }\textbf {\bibinfo {volume}
  {12}},\ \bibinfo {pages} {495--500} (\bibinfo {year} {1965})}\BibitemShut
  {NoStop}%
\bibitem [{\citenamefont {Ikeda}\ and\ \citenamefont
  {Matsumoto}(1987)}]{ikeda_high-dimensional_1987}%
  \BibitemOpen
  \bibfield  {author} {\bibinfo {author} {\bibfnamefont {K.}~\bibnamefont
  {Ikeda}}\ and\ \bibinfo {author} {\bibfnamefont {K.}~\bibnamefont
  {Matsumoto}},\ }\bibfield  {title} {\enquote {\bibinfo {title}
  {High-dimensional chaotic behavior in systems with time-delayed feedback},}\
  }\href {\doibase 10.1016/0167-2789(87)90058-3} {\bibfield  {journal}
  {\bibinfo  {journal} {Physica D}\ }\textbf {\bibinfo {volume} {29}},\
  \bibinfo {pages} {223--235} (\bibinfo {year} {1987})}\BibitemShut {NoStop}%
\bibitem [{\citenamefont {Shibata}\ and\ \citenamefont
  {Kaneko}(1998)}]{shibata_collective_1998}%
  \BibitemOpen
  \bibfield  {author} {\bibinfo {author} {\bibfnamefont {T.}~\bibnamefont
  {Shibata}}\ and\ \bibinfo {author} {\bibfnamefont {K.}~\bibnamefont
  {Kaneko}},\ }\bibfield  {title} {\enquote {\bibinfo {title} {Collective
  {Chaos}},}\ }\href {\doibase 10.1103/PhysRevLett.81.4116} {\bibfield
  {journal} {\bibinfo  {journal} {Phys. Rev. Lett.}\ }\textbf {\bibinfo
  {volume} {81}},\ \bibinfo {pages} {4116--4119} (\bibinfo {year}
  {1998})}\BibitemShut {NoStop}%
\bibitem [{\citenamefont {Cencini}\ \emph {et~al.}(1999)\citenamefont
  {Cencini}, \citenamefont {Falcioni}, \citenamefont {Vergni},\ and\
  \citenamefont {Vulpiani}}]{cencini_macroscopic_1999}%
  \BibitemOpen
  \bibfield  {author} {\bibinfo {author} {\bibfnamefont {M.}~\bibnamefont
  {Cencini}}, \bibinfo {author} {\bibfnamefont {M.}~\bibnamefont {Falcioni}},
  \bibinfo {author} {\bibfnamefont {D.}~\bibnamefont {Vergni}}, \ and\ \bibinfo
  {author} {\bibfnamefont {A.}~\bibnamefont {Vulpiani}},\ }\bibfield  {title}
  {\enquote {\bibinfo {title} {Macroscopic chaos in globally coupled maps},}\
  }\href {\doibase 10.1016/S0167-2789(99)00015-9} {\bibfield  {journal}
  {\bibinfo  {journal} {Physica D}\ }\textbf {\bibinfo {volume} {130}},\
  \bibinfo {pages} {58--72} (\bibinfo {year} {1999})}\BibitemShut {NoStop}%
\bibitem [{\citenamefont {Pazó}\ and\ \citenamefont
  {Montbrió}(2016)}]{pazo_quasiperiodic_2016}%
  \BibitemOpen
  \bibfield  {author} {\bibinfo {author} {\bibfnamefont {D.}~\bibnamefont
  {Pazó}}\ and\ \bibinfo {author} {\bibfnamefont {E.}~\bibnamefont
  {Montbrió}},\ }\bibfield  {title} {\enquote {\bibinfo {title} {From
  {Quasiperiodic} {Partial} {Synchronization} to {Collective} {Chaos} in
  {Populations} of {Inhibitory} {Neurons} with {Delay}},}\ }\href {\doibase
  10.1103/PhysRevLett.116.238101} {\bibfield  {journal} {\bibinfo  {journal}
  {Phys. Rev. Lett.}\ }\textbf {\bibinfo {volume} {116}},\ \bibinfo {pages}
  {238101} (\bibinfo {year} {2016})}\BibitemShut {NoStop}%
\bibitem [{Note3()}]{Note3}%
  \BibitemOpen
  \bibinfo {note} {The analytical derivations were assisted by ChatGPT 5.5,
  where all derivations and resulting formulas were verified
  manually.}\BibitemShut {Stop}%
\bibitem [{\citenamefont {Belykh}, \citenamefont {Belykh},\ and\ \citenamefont
  {Hasler}(2004)}]{belykh_blinking_2004}%
  \BibitemOpen
  \bibfield  {author} {\bibinfo {author} {\bibfnamefont {I.~V.}\ \bibnamefont
  {Belykh}}, \bibinfo {author} {\bibfnamefont {V.~N.}\ \bibnamefont {Belykh}},
  \ and\ \bibinfo {author} {\bibfnamefont {M.}~\bibnamefont {Hasler}},\
  }\bibfield  {title} {\enquote {\bibinfo {title} {Blinking model and
  synchronization in small-world networks with a time-varying coupling},}\
  }\href {\doibase 10.1016/j.physd.2004.03.013} {\bibfield  {journal} {\bibinfo
   {journal} {Physica D}\ }\textbf {\bibinfo {volume} {195}},\ \bibinfo {pages}
  {188--206} (\bibinfo {year} {2004})}\BibitemShut {NoStop}%
\bibitem [{\citenamefont {Porfiri}\ \emph {et~al.}(2006)\citenamefont
  {Porfiri}, \citenamefont {Stilwell}, \citenamefont {Bollt},\ and\
  \citenamefont {Skufca}}]{porfiri_random_2006}%
  \BibitemOpen
  \bibfield  {author} {\bibinfo {author} {\bibfnamefont {M.}~\bibnamefont
  {Porfiri}}, \bibinfo {author} {\bibfnamefont {D.~J.}\ \bibnamefont
  {Stilwell}}, \bibinfo {author} {\bibfnamefont {E.~M.}\ \bibnamefont {Bollt}},
  \ and\ \bibinfo {author} {\bibfnamefont {J.~D.}\ \bibnamefont {Skufca}},\
  }\bibfield  {title} {\enquote {\bibinfo {title} {Random talk: {Random} walk
  and synchronizability in a moving neighborhood network},}\ }\href {\doibase
  10.1016/j.physd.2006.09.016} {\bibfield  {journal} {\bibinfo  {journal}
  {Physica D}\ }\bibinfo {series} {Dynamics on {Complex} {Networks} and
  {Applications}},\ \textbf {\bibinfo {volume} {224}},\ \bibinfo {pages}
  {102--113} (\bibinfo {year} {2006})}\BibitemShut {NoStop}%
\bibitem [{\citenamefont {Stilwell}, \citenamefont {Bollt},\ and\ \citenamefont
  {Roberson}(2006)}]{stilwell_sufficient_2006}%
  \BibitemOpen
  \bibfield  {author} {\bibinfo {author} {\bibfnamefont {D.~J.}\ \bibnamefont
  {Stilwell}}, \bibinfo {author} {\bibfnamefont {E.~M.}\ \bibnamefont {Bollt}},
  \ and\ \bibinfo {author} {\bibfnamefont {D.~G.}\ \bibnamefont {Roberson}},\
  }\bibfield  {title} {\enquote {\bibinfo {title} {Sufficient {Conditions} for
  {Fast} {Switching} {Synchronization} in {Time}-{Varying} {Network}
  {Topologies}},}\ }\href {\doibase 10.1137/050625229} {\bibfield  {journal}
  {\bibinfo  {journal} {SIAM J. Appl. Dyn. Syst.}\ }\textbf {\bibinfo {volume}
  {5}},\ \bibinfo {pages} {140--156} (\bibinfo {year} {2006})}\BibitemShut
  {NoStop}%
\bibitem [{\citenamefont {Pecora}\ and\ \citenamefont
  {Carroll}(1998)}]{pecora_master_1998}%
  \BibitemOpen
  \bibfield  {author} {\bibinfo {author} {\bibfnamefont {L.~M.}\ \bibnamefont
  {Pecora}}\ and\ \bibinfo {author} {\bibfnamefont {T.~L.}\ \bibnamefont
  {Carroll}},\ }\bibfield  {title} {\enquote {\bibinfo {title} {Master
  {Stability} {Functions} for {Synchronized} {Coupled} {Systems}},}\ }\href
  {\doibase 10.1103/PhysRevLett.80.2109} {\bibfield  {journal} {\bibinfo
  {journal} {Phys. Rev. Lett.}\ }\textbf {\bibinfo {volume} {80}},\ \bibinfo
  {pages} {2109--2112} (\bibinfo {year} {1998})}\BibitemShut {NoStop}%
\bibitem [{Note4()}]{Note4}%
  \BibitemOpen
  \bibinfo {note} {The time-averaged Laplacian $\protect \bar {\protect \bm
  {L}}_n$ is defined as $\protect \bar {L}_{ij,n}=[\tau (t_n)]^{-1}\DOTSI
  \intop \ilimits@ _{t_{n-1}}^{t_n} dt\protect \, L_{ij}(t)$ and $\protect
  \overline {\Delta \theta }_n=[\tau (t_n)]^{-1}\DOTSI \intop \ilimits@
  _{t_{n-1}}^{t_n} dt\protect \, \Delta \theta (t)$. While the eigenvalues of
  the instantaneous Laplacian $\protect \bm {L}(t)$ are real, we obtain complex
  eigenvalues for the $\protect \bar {\protect \bm {L}}_{n}$ with a maximum
  imaginary part of order $10^{-2}$ over the whole dataset. We approximate $\mu
  _{\alpha ,n}\approx \protect \text {Re}(\mu _{\alpha ,n})$ and consider the
  MSF $\Lambda (\sigma \mu )$ only for real arguments since deviations from
  $\Lambda (\sigma \mu +\imath \delta \mu )$, $\delta \mu \in
  [-10^{-2},10^{-2}]$ are not visible over the plot range in Fig.~\ref
  {fig:laplacian_gap_vs_sync} (top panel). Significant deviations occur near
  $\sigma \mu =1$ but $\Lambda $ remains negative in this region.}\BibitemShut
  {Stop}%
\bibitem [{\citenamefont {Chen}\ and\ \citenamefont
  {Zheng}(2024)}]{chen_persistent_2024}%
  \BibitemOpen
  \bibfield  {author} {\bibinfo {author} {\bibfnamefont {Z.}~\bibnamefont
  {Chen}}\ and\ \bibinfo {author} {\bibfnamefont {Y.}~\bibnamefont {Zheng}},\
  }\bibfield  {title} {\enquote {\bibinfo {title} {Persistent and responsive
  collective motion with adaptive time delay},}\ }\href {\doibase
  10.1126/sciadv.adk3914} {\bibfield  {journal} {\bibinfo  {journal} {Sci.
  Adv.}\ }\textbf {\bibinfo {volume} {10}},\ \bibinfo {pages} {eadk3914}
  (\bibinfo {year} {2024})}\BibitemShut {NoStop}%
\bibitem [{\citenamefont {Hart}\ \emph {et~al.}(2017)\citenamefont {Hart},
  \citenamefont {Schmadel}, \citenamefont {Murphy},\ and\ \citenamefont
  {Roy}}]{hart_experiments_2017}%
  \BibitemOpen
  \bibfield  {author} {\bibinfo {author} {\bibfnamefont {J.~D.}\ \bibnamefont
  {Hart}}, \bibinfo {author} {\bibfnamefont {D.~C.}\ \bibnamefont {Schmadel}},
  \bibinfo {author} {\bibfnamefont {T.~E.}\ \bibnamefont {Murphy}}, \ and\
  \bibinfo {author} {\bibfnamefont {R.}~\bibnamefont {Roy}},\ }\bibfield
  {title} {\enquote {\bibinfo {title} {Experiments with arbitrary networks in
  time-multiplexed delay systems},}\ }\href {\doibase 10.1063/1.5016047}
  {\bibfield  {journal} {\bibinfo  {journal} {Chaos: An Interdisciplinary
  Journal of Nonlinear Science}\ }\textbf {\bibinfo {volume} {27}},\ \bibinfo
  {pages} {121103} (\bibinfo {year} {2017})}\BibitemShut {NoStop}%
\bibitem [{\citenamefont {Hart}\ \emph
  {et~al.}(2019{\natexlab{b}})\citenamefont {Hart}, \citenamefont {Larger},
  \citenamefont {Murphy},\ and\ \citenamefont {Roy}}]{hart_delayed_2019}%
  \BibitemOpen
  \bibfield  {author} {\bibinfo {author} {\bibfnamefont {J.~D.}\ \bibnamefont
  {Hart}}, \bibinfo {author} {\bibfnamefont {L.}~\bibnamefont {Larger}},
  \bibinfo {author} {\bibfnamefont {T.~E.}\ \bibnamefont {Murphy}}, \ and\
  \bibinfo {author} {\bibfnamefont {R.}~\bibnamefont {Roy}},\ }\bibfield
  {title} {\enquote {\bibinfo {title} {Delayed dynamical systems: networks,
  chimeras and reservoir computing},}\ }\href {\doibase 10.1098/rsta.2018.0123}
  {\bibfield  {journal} {\bibinfo  {journal} {Phil. Trans. R. Soc. A}\ }\textbf
  {\bibinfo {volume} {377}},\ \bibinfo {pages} {20180123} (\bibinfo {year}
  {2019}{\natexlab{b}})}\BibitemShut {NoStop}%
\bibitem [{\citenamefont {Müller-Bender}\ \emph {et~al.}(2020)\citenamefont
  {Müller-Bender}, \citenamefont {Otto}, \citenamefont {Radons}, \citenamefont
  {Hart},\ and\ \citenamefont {Roy}}]{muller-bender_laminar_2020}%
  \BibitemOpen
  \bibfield  {author} {\bibinfo {author} {\bibfnamefont {D.}~\bibnamefont
  {Müller-Bender}}, \bibinfo {author} {\bibfnamefont {A.}~\bibnamefont
  {Otto}}, \bibinfo {author} {\bibfnamefont {G.}~\bibnamefont {Radons}},
  \bibinfo {author} {\bibfnamefont {J.~D.}\ \bibnamefont {Hart}}, \ and\
  \bibinfo {author} {\bibfnamefont {R.}~\bibnamefont {Roy}},\ }\bibfield
  {title} {\enquote {\bibinfo {title} {Laminar chaos in experiments and
  nonlinear delayed {Langevin} equations: {A} time series analysis toolbox for
  the detection of laminar chaos},}\ }\href {\doibase
  10.1103/PhysRevE.101.032213} {\bibfield  {journal} {\bibinfo  {journal}
  {Phys. Rev. E}\ }\textbf {\bibinfo {volume} {101}},\ \bibinfo {pages}
  {032213} (\bibinfo {year} {2020})}\BibitemShut {NoStop}%
\bibitem [{\citenamefont {Santra}, \citenamefont {Basu},\ and\ \citenamefont
  {Sabhapandit}(2020)}]{santra_run-and-tumble_2020}%
  \BibitemOpen
  \bibfield  {author} {\bibinfo {author} {\bibfnamefont {I.}~\bibnamefont
  {Santra}}, \bibinfo {author} {\bibfnamefont {U.}~\bibnamefont {Basu}}, \ and\
  \bibinfo {author} {\bibfnamefont {S.}~\bibnamefont {Sabhapandit}},\
  }\bibfield  {title} {\enquote {\bibinfo {title} {Run-and-tumble particles in
  two dimensions: {Marginal} position distributions},}\ }\href {\doibase
  10.1103/PhysRevE.101.062120} {\bibfield  {journal} {\bibinfo  {journal}
  {Phys. Rev. E}\ }\textbf {\bibinfo {volume} {101}},\ \bibinfo {pages}
  {062120} (\bibinfo {year} {2020})}\BibitemShut {NoStop}%
\bibitem [{\citenamefont {Bellen}\ and\ \citenamefont
  {Zennaro}(2003)}]{bellen_numerical_2003}%
  \BibitemOpen
  \bibfield  {author} {\bibinfo {author} {\bibfnamefont {A.}~\bibnamefont
  {Bellen}}\ and\ \bibinfo {author} {\bibfnamefont {M.}~\bibnamefont
  {Zennaro}},\ }\href {\doibase 10.1093/acprof:oso/9780198506546.001.0001}
  {\emph {\bibinfo {title} {Numerical {Methods} for {Delay} {Differential}
  {Equations}}}}\ (\bibinfo  {publisher} {Oxford University Press},\ \bibinfo
  {address} {New York},\ \bibinfo {year} {2003})\BibitemShut {NoStop}%
\end{thebibliography}%

\end{document}